\documentclass[twocolumn]{aastex63}
\usepackage{hyperref}
\usepackage{rotating}

\shortauthors{Foo et al.}
\graphicspath{{./}{figures/}}

\begin{document}

\title{PASSAGES: The Discovery of a Strongly Lensed Protocluster Core Candidate at Cosmic Noon}

\correspondingauthor{Nicholas Foo}
\email{nfoo1@asu.edu}

\author[0000-0002-7460-8460]{Nicholas Foo}
\affiliation{School of Earth \& Space Exploration, Arizona State University, Tempe, AZ 85287-1404, USA}
\affiliation{Department of Astronomy/Steward Observatory, University of Arizona, 933 N. Cherry Avenue, Tucson, AZ 85721, USA}

\author[0000-0001-5429-5762]{Kevin C. Harrington}
\affiliation{Joint ALMA Observatory, Alonso de C{\'o}rdova 3107, Vitacura, Casilla 19001, Santiago de Chile, Chile}
\affiliation{National Astronomical Observatory of Japan,
Los Abedules 3085 Oficina 701, Vitacura 763 0414, Santiago, Chile}
\affiliation{European Southern Observatory, Alonso de C{\'o}rdova 3107, Vitacura, Casilla 19001, Santiago de Chile, Chile}
\affiliation{Instituto de Estudios Astrofísicos, Facultad de Ingeniería y 455 Ciencias, Universidad Diego Portales, Av. Ejército Libertador 441, Santiago, Chile}

\author[0000-0003-1625-8009]{Brenda L. Frye}
\affiliation{Department of Astronomy/Steward Observatory, University of Arizona, 933 N. Cherry Avenue, Tucson, AZ 85721, USA}

\author[0000-0001-9394-6732]{Patrick S. Kamieneski}
\affiliation{School of Earth \& Space Exploration, Arizona State University, Tempe, AZ 85287-1404, USA}

\author[0000-0001-7095-7543]{Min S. Yun}
\affiliation{Department of Astronomy, University of Massachusetts, Amherst, MA 01003, USA}

\author[0000-0002-2282-8795]{Massimo Pascale}
\affiliation{Department of Astronomy, University of California, 501 Campbell Hall \#3411, Berkeley, CA 94720, USA}

\author[0000-0001-9163-0064]{Ilsang Yoon}
\affiliation{National Radio Astronomy Observatory, 520 Edgemont Road, Charlottesville, VA 22903}

\author[0000-0003-1832-4137]{Allison Noble}
\affiliation{School of Earth \& Space Exploration, Arizona State University, Tempe, AZ 85287-1404, USA}

\author[0000-0001-8156-6281]{Rogier A. Windhorst}
\affiliation{School of Earth \& Space Exploration, Arizona State University, Tempe, AZ 85287-1404, USA}

\author[0000-0003-3329-1337]{Seth H. Cohen}
\affiliation{School of Earth \& Space Exploration, Arizona State University, Tempe, AZ 85287-1404, USA}

\author[0000-0001-9969-3115]{James D. Lowenthal}
\affiliation{Smith College, Northampton, MA 01063, USA}

\author[0000-0002-1173-2579
]{Melanie Kaasinen}
\affiliation{European Southern Observatory, Karl-Schwarzschild-Strasse 2, D-85748 Garching, Germany}

\author[0000-0002-4140-0428]{Bel\'{e}n Alcalde Pampliega}
\affiliation{European Southern Observatory, Alonso de C{\'o}rdova 3107, Vitacura, Casilla 19001, Santiago de Chile, Chile}

\author[0000-0001-9773-7479
]{Daizhong Liu}
\affiliation{Purple Mountain Observatory, Chinese Academy of Sciences, 10
Yuanhua Road, Nanjing 210023, China}

\author[0000-0003-3881-1397]{Olivia Cooper}
\affiliation{The University of Texas at Austin, 2515 Speedway Boulevard Stop C1400, Austin, TX 78712, USA}

\author[0000-0002-4223-2016
]{Carlos Garcia Diaz}
\affiliation{Department of Astronomy, University of Massachusetts, Amherst, MA 01003, USA}

\author[0000-0003-0748-4768
]{Anastasio Díaz-Sánchez}
\affiliation{Departamento Física Aplicada, Universidad Politécnica de Cartagena, Campus Muralla del Mar, 30202 Cartagena, Murcia, Spain}

\author[0000-0001-9065-3926
]{Jose Diego}
\affiliation{FCA, Instituto de Fisica de Cantabria (UC-CSIC), Av.  de Los Castros s/n, E-39005 Santander, Spain}

\author[0000-0003-3418-2482]{Nikhil Garuda}
\affiliation{Department of Astronomy/Steward Observatory, University of Arizona, 933 N. Cherry Avenue, Tucson, AZ 85721, USA}

\author[0000-0002-2640-5917]{Eric F. Jim\'{e}nez-Andrade}
\affiliation{Instituto de Radioastronom\'{i}a y Astrof\'{i}sica, Universidad Nacional Aut\'{o}noma de M\'{e}xico, Antigua Carretera a P\'{a}tzcuaro \# 8701, Ex-Hda. San Jos\'{e} de la Huerta, Morelia, Michoac\'{a}n, C.P. 58089, M\'{e}xico}

\author[0009-0001-7446-2350]{Reagen Leimbach}
\affiliation{Department of Astronomy/Steward Observatory, University of Arizona, 933 N. Cherry Avenue, Tucson, AZ 85721, USA}

\author[0000-0002-4444-8929]{Amit Vishwas}
\affiliation{Cornell Center for Astrophysics and Planetary Science, Cornell University, Space Sciences Building, Ithaca, NY 14853, USA}

\author[0000-0002-9279-4041]{Q. Daniel Wang}
\affiliation{Department of Astronomy, University of Massachusetts, Amherst, MA 01003, USA}

\author[0000-0002-6922-469X
]{Dazhi Zhou}
\affiliation{Department of Physics and Astronomy, University of British Columbia, 6225 Agricultural Rd., Vancouver, V6T 1Z1, Canada}

\author[0000-0002-0350-4488
]{Adi Zitrin}
\affiliation{Department of Physics, Ben-Gurion University of the Negev, P.O. Box 653, Be’er-Sheva 84105, Israel}


\begin{abstract}
Investigating the processes by which galaxies rapidly build up their stellar mass during the peak of their star formation ($z=2-3$) is crucial to advancing our understanding of the assembly of large-scale structures. We report the discovery of one of the most gas- and dust-rich protocluster core candidates, 
PJ0846+15 (J0846), from the Planck All-Sky Survey to Analyze Gravitationally lensed Extreme Starbursts (PASSAGES) sample. The exceedingly high total star formation rate (SFR) uncorrected for lensing magnification ($\mu$) of $\mu$SFR $=39900^{+23000}_{-12900}~\textup{ M}_\odot$~yr$^{-1}$ is the result of a foreground cluster lensing at least 11 dusty star-forming galaxies between $z=2.660-2.669$, where the intrinsic value is estimated to be SFR $=5200^{+3200}_{-2000}~\textup{ M}_\odot$ yr$^{-1}$. Atacama Large Millimeter Array (ALMA) observations uncovered 18 CO(3–2) emission-line detections, some of which are multiply-imaged systems, lensed by a foreground cluster at $z=0.77$. 
We present the first multi-wavelength characterization of this field, constructing a lens model that predicts that these 11 galaxies
($\mu\simeq1.5-25$) are contained within a projected physical extent of 280 $\times$ 150 kpc, with a velocity dispersion of $\sigma_{v}=246\pm72$ km s$^{-1}$. J0846 exhibits the rare case of a protocluster candidate whose core is strongly-lensed, offering a magnified view of the rapid stellar buildup within an overdense environment at Cosmic Noon.

\end{abstract}

\keywords{Protoclusters (1297); Strong gravitational lensing (1643); Starburst galaxies (1570); Galaxy evolution (594); Interstellar medium (847) }

\section{Introduction} \label{sec:intro}

Studying the progenitors of massive galaxy clusters, protoclusters \cite[see][]{Overzier_2016}, provides important insights into the assembly of large-scale structures across cosmic time. 
During the empirically determined peak of the co-moving star-formation rate density (SFRD), at $z\simeq1-3$ \citep{Madau2014}, it is expected that member galaxies of protoclusters contributed $20-50$\% towards the SFRD \citep{Chiang_2017}. Subsequently, mature clusters identified at $z<1$ are often dominated by massive quiescent ellipticals \citep{Dressler_1980} with an established red sequence \citep{Gladders_2000}, especially within their cluster cores \citep{Papovich_2010, Stanford_2012}. The physical processes responsible for driving the transition of star formation (SF) activity remain poorly understood. Constraining such mechanisms is crucial to linking galaxy evolution and cosmological models to reproduce and explain the intense star formation rate (SFR) and subsequent quenching observed in protoclusters at Cosmic Noon \citep{Hill_2020, Lim_2021}. 
\par
Various methods have been employed to discover protoclusters, including searches for Lyman $\alpha$ emitters \citep[LAEs, e.g.,][]{Venemans_2007, Chiang_2015}, Lyman Break Galaxies \cite[LBGs; e.g.,][]{Steidel_1998, Toshikawa_2018}, H$\alpha$ emitters \cite[HAEs; e.g.,][]{Hatch_2011, Koyama_2013}, active galactic nuclei \cite[AGN; e.g.,][]{Wylezalek2013, Hatch_2014, chapman2023} and sub-millimeter (sub-mm) bright dusty star-forming galaxies \cite[DSFGs; e.g.,][]{Casey_2016, Chapman_2009, Dannerbauer_2014}. In several cases they even trace the same overdense structures \citep[e.g.,][]{Umehata_2015, Harikane_2019,Champagne_2021}, suggesting that extragalactic protoclusters consists of a diverse populations of galaxies with a wide range of physical properties leading to different evolutionary paths of star formation history. In particular, sub-mm selected, DSFG-rich protocluster candidates likely represent the most active star-bursting regions during cluster formation. Such DSFG-rich systems have often been found to be embedded within protocluster cores,  with total SFR, in the order of thousands of $\textup{ M}_\odot$ yr$^{-1}$ \citep{ Cucciati_2018, Miller_2018, Oteo_2018,Champagne_2021}. 
These active cores at $z>2$ may harbor the progenitors that will form the massive ellipticals and even the central brightest cluster galaxies (BCGs) that we observe in local clusters.

To date, only a handful of strongly lensed galaxy protocluster candidates have been reported. \cite{Ishigaki2016, Morishita_2023} and \cite{Laporte_2022} both identified $z>7$ protoclusters behind the galaxy clusters of Abell 2744 and SMACS0723-7327, respectively, but neither exhibited evidence of strong lensing (producing multiply-imaged systems). A galaxy group consisting of $>4$ members at $z=4.32$ has been reported to be multiply-imaged by the massive \textit{``El Gordo}"  foreground cluster  \citep{Caputi_2021, Frye_2023}. One of the few sub-mm selected examples reported to date is a group of $>4$ moderately star-forming (total SFR$=450\pm50\textup{ M}_\odot$ yr$^{-1}$) galaxies  at $z=2.9$ with CO detections from ALMA, strongly-lensed by the MS 0451.6-0305 foreground cluster \citep{Borys_2004,MacKenzie_2014, Shen_2021}.



\par
Sub-mm surveys were undertaken by \textit{\textit{Planck}} and Herschel which provided all-sky and wide-field maps to search for bright (e.g., strongly lensed) DSFGs and protoclusters \citep{PlanckCollaboration2015, Canameras_2015, Harrington2016, Diaz_2017}. To identify the brightest high-redshift DSFGs in the \textit{Planck} all-sky survey, \cite{Harrington2016} and \cite{Berman2022} carried out cross-match identifications with \textit{Herschel} and the Wide-field Infrared Survey Explorer (\textit{WISE}) applying a color-magnitude filtering process removing foreground Galatic cirrus sources and radio-loud AGN \citep{Yun2008}. A sample of  $\sim$30 of the brightest objects in the sky were identified to be DSFGs gravitationally lensed by a mix of both foreground galaxies or clusters/groups of galaxies. This prompted follow-up spectroscopic observations conducted to confirm the redshifts of these sources, ranging from $z=1.1-3.5$ \citep{Harrington2016, Berman2022}, characterizing this parent sample of the \textit{\textit{Planck}} All-Sky Survey to Analyze Gravitationally-lensed Extreme Starbursts (PASSAGES)\footnote{\url{https://sites.google.com/view/astropassages}}. \cite{Harrington2021} conducted a single-dish campaign with the Large Millimeter Telescope (LMT), Institute for Radio Astronomy in the Millimeter Range (IRAM), and Green Bank Telescope (GBT) detecting $>160$ CO and [CI] emission lines and performed a full line and dust continuum radiative transfer analysis of the global gas excitation properties. Optical/near-IR imaging was obtained with Hubble Space Telescope (\textit{HST}) and Gemini to investigate the foreground lenses and their associated strong lensing properties \citep[][Lowenthal et al. in prep]{Kamieneski_2024_PASSAGES}, leading to the following discovery.
\par
One of these PASSAGES sources, PJ0846+15 (J0846), is found to be lensed by a foreground galaxy cluster determined to be at $z=0.77$. Here we present the discovery that J0846 does not consist of just one source, but instead at least $11$ CO(3–2) emitting galaxies all at $z = 2.67 $ uncovered by Atacama Large Millimeter Array (ALMA) observations. Four of these sources are strongly-lensed into multiple images, or image systems. To distinguish the foreground cluster lens at $z=0.77$ and the background lensed system at $z=2.67$, hereafter we refer to them as J0846.FG and J0846.BG, respectively.\\

\par
This manuscript is organized as follows. In Section \ref{Sec:Observations} we describe the multi-wavelength observations from the ALMA, \textit{\textit{HST}}, Gemini, Very Large Array (VLA), with optical spectroscopy from the Very Large Telescope (VLT) using the Multi-Unit Spectroscopic Explorer (MUSE) and the VLT FOcal Reducer/low dispersion Spectrograph 2 (FORS2), as well as the Gemini Multi-Object Spectrograph (Gemini/GMOS). We show the results of ALMA CO(3–2) detections in Section \ref{Sec:The CO Detections at $z=2.67$}. 
In Section \ref{Sec:The Lens Model}, we present the lens model of the foreground cluster (J0846.FG). In Section \ref{Sec:Global Properties of a Protocluster Candidate} we investigate the properties of J0846.BG and its 11 members. In Section \ref{Sec:Discussion}, we begin to discuss J0846.BG as a protocluster core and in the context of cluster formation. Section \ref{Sec:Conclusion} presents our conclusions as well as the outlook moving forward. This paper assumes a flat $\Lambda$CDM cosmology with $H_0=70$ km s$^{-1}$ Mpc$^{-1}$, $\Omega_{m}=0.3$, $\Omega_{\Lambda}=0.7$ and 1$^{\prime\prime}$=8.107 kpc at $z=2.67$.

\section{Multi-Wavelength Observations} \label{Sec:Observations}
Here we provide details for multi-wavelength imaging and spectroscopy to determine the spectroscopic redshifts and properties of both the foreground cluster lens (J0846.FG) and the background lensed objects at $z=2.67$ (J0846.BG).




\subsection{ALMA}
ALMA observations were acquired in a Band 3  Cycle 5 program (2017.1.01214.S, PI: Min Yun). These observations were taken using the 12 m antennae array in configuration 3 with baselines ranging from $15-2517$ m, and an average on-source integration time of 4596.480 seconds. Observations were set up with two overlapping pointings, with a primary beam of $\sim60^{\prime\prime}$, targeting the two \textit{Planck}-\textit{WISE} selected objects with LMT/RSR CO(3–2) detections reported in \cite{Berman2022}: PJ084650.1 and PJ084648.6. The total frequency bandwidth spans $91.53-107.151$ GHz with a resolution of 31.25 MHz for the continuum, while one sub-band was set up to resolve the CO(3–2) line emission with a resolution of 3.9043 MHz, equivalent to 
12.279 km s$^{-1}$. We reduced the dataset with the Common Astronomy Software Application \citep[CASA,][]{McMullin2007, CASA_2022} version 6.5.4.9 with the ALMA pipeline heuristics \citep{Hunter_2023}, using the \texttt{tclean} routine to produce both spectral line data cubes and continuum images. We utilized the \texttt{uvcontsub} routine in CASA to subtract the observed 3mm continuum, ensuring that only line-free channels were included for relevant analyses. 

\par
\begin{figure*}
	\centering\includegraphics[scale =0.272]{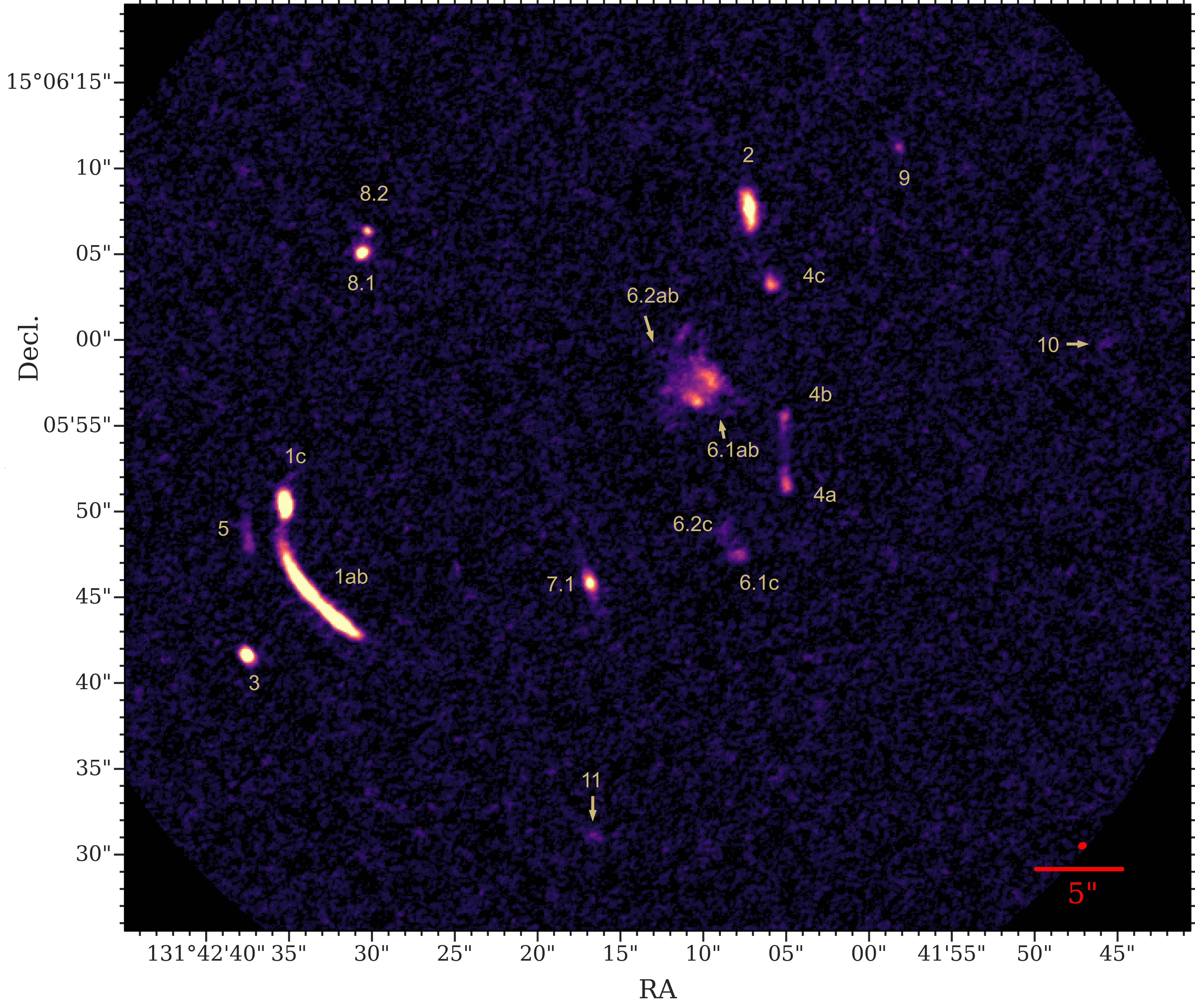}
	\caption{The ALMA CO(3–2) line integrated moment-0 map of J0846.BG consisting of 18 lensed detections at $z=2.660-2.669$ all within 60$^{\prime\prime}$ of each other. Their associated 1-D spectra for each detection is displayed in Figure \ref{fig:2}. Each detection is labeled based off their image system designation (see Section \ref{Sec:The Image Systems}). In total these 18 CO images originate from 11 unique sources. The red ellipse in the bottom right represents the synthesized beam.  }
	\label{fig:1}
\end{figure*}

Using the \texttt{tclean} task in the CASA reduction pipeline both the CO(3–2) line and continuum imaging were generated with a pixel size of $0.09^{\prime\prime}$ and cleaned using Briggs weighting with a robust parameter of 1.0. This resulted in a synthesized beam size of $0.56^{\prime\prime}\times0.46^{\prime\prime}$, a position angle of $-63.18 ^\circ$. To produce the image cubes, we opt for a channel width of 15.625MHz ($\sim$50 km s$^{-1}$), a factor of four greater than the native resolution, to increase the S/N of potentially faint detections. We also generate versions of the image cube using a finer channel width of 25 km s$^{-1}$ to further inspect any additional velocity components within each CO(3–2) line profile. The final root mean square (rms) achieved for the continuum-subtracted line map was $\sim0.1$ mJy beam$^{-1}$ per 50 km s$^{-1}$ channel. We also generate intensity-weighted velocity (moment 1) maps, collapsing the velocity axis weighted by the intensity of each pixel for each source using a customized velocity range and aperture. The 3 mm continuum image we produced included multi-frequency synthesis of all line-free channels from the Band 3 observations, utilizing all spectral windows covering $91.53-107.151$ GHz. The final rms achieved for the continuum imaging was $\sim6$ $\mu$Jy beam$^{-1}$. To measure continuum flux values, apertures were based on 3$\sigma$ threshold contours, utilizing the same procedure as the CO(3–2) line extraction. We find that only 8/18 CO detections have 3 mm continuum counterparts (see Section \ref{Sec:The CO Detections at $z=2.67$}). We show the CO(3-2) moment 0 map in Figure \ref{fig:1}.




\begin{figure*}
	\centering\includegraphics[scale =1.55]{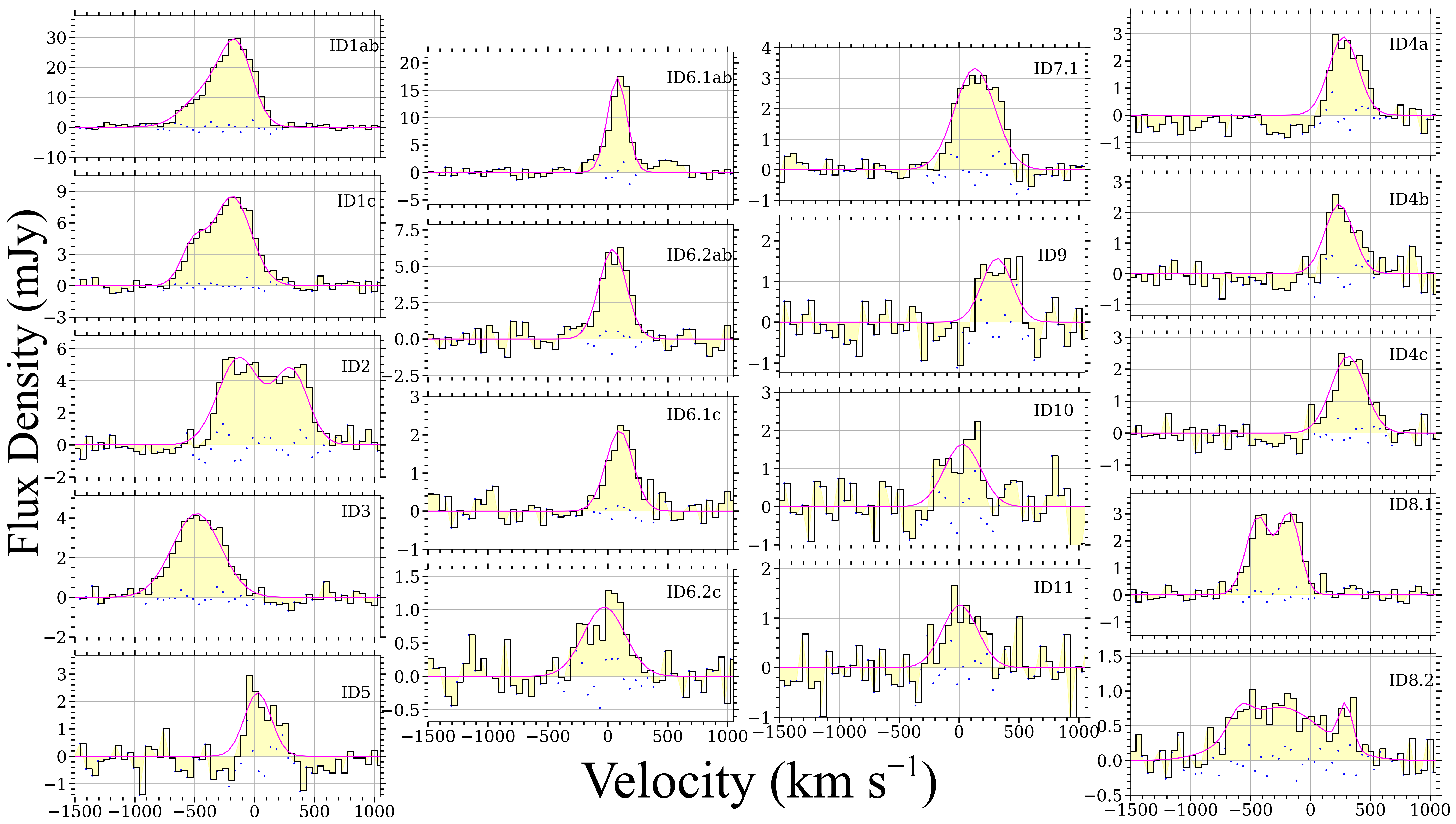}
	\caption{ALMA 1-D spectra for all 18 CO(3–2) line detections with their best-fit multi-Gaussian profiles plotted in the magenta curve (residual is plotted in the blue points). All detections are within the redshift range of $z=2.660-2.669$, equating to $\Delta V\approx800$ km s$^{-1}$ of each other. The central velocity, $\Delta V=0$ km s$^{-1}$, corresponds to $z=2.666$, the central redshift of this system.}  
	\label{fig:2}
\end{figure*}

\subsection{VLA 6 GHz Continuum}
Karl G. Jansky VLA 6 GHz observations were acquired as part of a larger program (18A-399, PI: P. Kamieneski). The observational details are  presented in \cite{Kamieneski_2024_PASSAGES} along with the full data reduction procedure. Briefly, observations were carried out at 5 cm with the C band receivers (4–8 GHz)
and full polarization for 1.5 hours utilizing the WIDAR correlator configured to 3-bit sampling. The effective combined bandwidth of the two basebands is 4 GHz, centered at 6 GHz. We use the VLA data to provide an initial morphological characterization of J0846.FG and determine radio counterparts to the J0846.BG members. 
\par
The 6 GHz continuum image is produced using the CASA reduction pipeline with the \texttt{tclean} task. We apply a Briggs weighting with robust parameter of 0.5, and a pixel size of $0.05^{\prime\prime}$ resulting in a synthesized beam size of $0.37^{\prime\prime} \times0.29^{\prime\prime}$ at a position angle of $60.41^\circ$. The continuum noise rms level achieved is 3.6 $\mu$Jy beam$^{-1}$.


\subsection{\textit{HST} Observations}
Imaging from the Wide Field Camera 3 IR detector (WFC3-IR) aboard \textit{HST} was obtained as part of a larger Cycle 24 program to confirm the strongly lensed nature for the candidates within the sample of PASSAGES targets (GO-14653; PI: J. Lowenthal). We refer to \cite{Kamieneski_2024_PASSAGES} for a review. A single orbit was taken with the F160W wide-band filter, composed of a five-point dithering pattern, of 500~s each, resulting in a subpixel dithering setup of 0.572$^{\prime\prime}$ spacing. The individual exposures were pipeline-reduced, flat-fielded, and calibrated with standard Space Telescope Science Institute (STScI) routines. The exposures were then combined using astrodrizzle (\texttt{final\_pixfrac} of 0.9). We adopt a native pixel scale to reach 0.065$^{\prime\prime}$, with a point-spread function (PSF) Full Width Half Maximum (FWHM) of 0.2$^{\prime\prime}$. The final image has a FOV$= 141^{\prime\prime} \times 125^{\prime\prime}$ and reaches a sensitivity ($5\sigma$) of $m_{AB} \sim 28.7$ for an unresolved point source. Further details on the photometry can be found in Appendix \ref{Appendix:Optical-Infrared Imaging}.

\subsection{Gemini Observations}
Gemini/GMOS r$^{\prime}$ and z$^{\prime}$ imaging and slit spectroscopy was acquired as part of the Gemini South programs GS-2018A-Q-216, GS-2018B-Q-123, GS-2020A-Q-217  (PI: James Lowenthal). The program was executed on 2018 July 13–14 for semester 2018A, on 2018 October 7 for semester 2018B, and on 2020 February 20 for semester 2020A. This data was part of a larger program that was described in \cite{Kamieneski_2024_PASSAGES}. Imaging was obtained with the r$^{\prime}$ filter for 1500 s to achieve a signal-to-noise ratio (S/N) = 10 for an object with r$^{\prime}$ = 25.0 mag, and in the z$^{\prime}$ filter for 2100 s to achieve an S/N = 10 for a z$^{\prime}$ = 23.0 mag object. Observations were taken with dithered 300 s exposures (five positions for r$^{\prime}$ and seven positions for z$^{\prime}$) to cover chip gaps within the GMOS 5.5$^{\prime}$ $\times$ 5.5$^{\prime}$ field of view (FOV), resulting in a pixel scale of 0.080$^{\prime\prime}$. Additional details on the photometry can be found in Appendix \ref{Appendix:Optical-Infrared Imaging}. 

\subsection{VLT Observations}

Very Large Telescope observations were obtained with the FOcal Reducer and low dispersion Spectrograph 2 (FORS2) and Multi Unit Spectroscopic Explorer (MUSE) instruments with the aim of measuring spectroscopic redshifts for the foreground cluster members and any line-of-sight structure, as well as background lensed systems. Owing to unfavorable observing conditions in these filler programs, we were unable to make detections of any lensed background sources, only managing to secure redshifts of foreground objects. We provide further details of the optical spectroscopy in Appendix \ref{Appendix:Optical Spectroscopy}.

\subsubsection{FORS2}
A VLT UT1 filler time program (PI C. Mazzuchelli, PID:110.23UC) was executed in seven observing blocks using the FORS2 MOS (multi-object spectroscopy) mode between February 19 and March 8 2023, across four different nights. There were four masks, each consisting of six or seven 1.2$^{\prime\prime}$ slits placed in the central FOV of J0846, each observed 1-2 times. Each block began with a 60 second acquisition image using the red imaging filter, R Special76, followed by a 60 second slit acquisition (`FORS2-mos-obs-slit'), before acquiring two 1300s exposures using the `FORS2-mos-obs-off' observational setup and filter/grism pair: GG435+81/GRIS-600RI+19. Seeing conditions varied about a range of between $0.8-1.1^{\prime\prime}$, measuring off the FWHM of skylines in the 2D spectra. Data reduction was performed utilizing the FORS2 pipeline version 2.8.5 \citep{Weilbacher_2020} based in the \texttt{EsoReflex} workflow environment \citep{Freudling_2013}. This includes all standard bias subtraction, flat-fielding, dark-field correction, wavelength calibration, sky subtraction and flux calibrations generating the science ready data products. The final 1D spectra provides coverage in the range of $512-849$ nm. We remove any cosmic rays that appear in the extracted 1-D spectra by comparing multiple exposures for each extracted source object by a sigma-clip in a custom-built code.

%

\subsubsection{MUSE}
Integral field spectrograph (IFU) MUSE observations were obtained as part of an ESO filler time program (PI F. Bian, PID:111.24UJ.006). This was carried out utilizing the wild field mode without the capabilities of adaptive optics (WFM-NOAO-N), resulting in a seeing range of $0.56 - 1.74^{\prime\prime}$.  The observations were executed on May 4, 2023, with a total integration time of 2480s over four exposures, and with a single pointing providing a 1$^{\prime}$ square FOV that only partially covers the full extent of the cluster field. Data reduction was performed utilizing the standard MUSE pipeline version 2.8.5 \citep{Weilbacher_2020} within the \texttt{EsoReflex} \citep{Freudling_2013} workflow environment. This applies all the required corrections and calibrations including bias subtraction, flat-fielding, dark field correction, wavelength calibration, sky subtraction and flux calibrations. We used the Zurich Atmosphere Purge (ZAP) tool software \citep{Soto2016}, to further subtract any skyline residuals missed by the initial standard reduction method. The final 3D datacube provides coverage in the spectral range of $475 - 935$ nm.\\

\vspace{1cm}





\begin{deluxetable}{cccc}[h] \label{Table:1}
\tablecaption{ALMA CO(3–2) Detections at $z=2.67$}
\tablecolumns{12}
\tablewidth{0pc}
\tablehead{
\colhead{ID} &  \colhead{R.~A.} & \colhead{Decl.} & \colhead{$z_{spec}$}\\
\colhead{}   & \colhead{(J2000)}      & \colhead{(J2000)}      & \colhead{} 
}
\startdata
\ 
1ab &08:46:50.2598& +15:05:45.508 &2.663\\ 
1c & 08:46:50.3487 & +15:05:50.436 &2.663\\
2& 08:46:48.4811 & +15:06:07.685 &2.665\\ 
3 &08:46:50.5051 &+15:05:41.622  &2.660\\ 
4a & 08:46:48.3329&+15:05:51.518  &2.668 \\ 
4b & 08:46:48.3379& +15:05:55.496 &2.668\\ 
4c & 08:46:48.3903 &+15:06:03.200  &2.669\\ 
5 & 08:46:50.5038 & +15:05:48.673 &2.666\\ 
6.1ab & 08:46:48.6466& +15:05:57.572 &2.666\\ 
6.1c & 08:46:48.4979 & +15:05:47.241 &2.667\\
6.2ab &  08:46:48.7566& +15:06:00.030&2.666\\ 
6.2c & 08:46:48.5763 & +15:05:48.943 &2.665\\
7 & 08:46:49.1244 & +15:05:45.795 &2.666\\ 
8.1 & 08:46:50.0357 & +15:06:05.044 &2.661\\ 
8.2 & 08:46:50.0184& +15:06:06.341 &2.662\\ 
9 & 08:46:47.8800 &+15:06:11.254&2.669\\ 
10 &08:46:47.0457 & +15:05:59.734 &2.668\\ 
11 & 08:46:49.1044 &+15:05:31.137 &2.664\\ 
\hline
\enddata
\tablecomments{ Column 1: ID; Column 2: R.~A.; Column 3: Decl.; Column 4: CO(3–2) spectroscopic redshift}


  \label{tab_GGs}
\end{deluxetable}

\section{The CO Detections at $z=2.67$} \label{Sec:The CO Detections at $z=2.67$}





Based on the two \textit{Planck}-\textit{WISE} selected candidates reported with CO(3–2) line detections at $z=2.664$ and $z=2.661$ in \cite{Berman2022}, we searched for individual detections in the generated ALMA image-cubes prior to primary-beam corrections. We search for any counterparts in F160W, r$^{\prime}$ and z$^{\prime}$ to further corroborate secure detections. After an intial by-eye inspection of the cubes we subsequently conduct a blind search for CO(3–2) emission in the data cube\footnote{Individual execution blocks (EB) were imaged and inspected independently to verify the fidelity of line features.} utilizing the \texttt{FindClumps} algorithm \citep{Walter_2016} built into the \texttt{Interferopy} python package \citep{interferopy}. To substantiate a secure line detection we impose a criterion of S/N$>2.5$ per channel, across $\geq3$ continuous channels. In total we report 18 CO(3–2) detections (Table \ref{Table:1}). We also serendipitously discovered a line detection in one of the side-bands at 92.737 GHz. Assuming that this line corresponds to CO(3–2) would place it $\sim5000$ km s$^{-1}$ offset from the other 18 CO(3-2) emission line detections from the lensed J0846.BG. For this work, we have not included this detection in our analyses as it unlikely to be a protocluster member.
\par
To extract the 1-D spectra of each source we generated individual moment-0 maps to measure the CO(3–2) line detections (Figure \ref{fig:1}). In many instances the morphology of the CO detections become distorted and complex due to lensing effects. Therefore, based on their moment-0 maps, apertures were manually placed to encompass all pixels above a $>3\sigma$ threshold. We model the spectral line profile utilizing multi-Gaussian functions and integrate them to obtain their line fluxes, measuring from the primary beam corrected image cubes (Figure \ref{fig:2}). Following the procedure described in, e.g., \cite{Solomon2005}, we calculate the CO(3–2) line luminosities. We provide the measured properties of the 18 CO(3–2) detection in Table \ref{Table:2}.

\begin{deluxetable*}{ccccccccc}
\label{Table:2}
\tablecaption{Apparent Derived Properties for CO(3–2) detections at $z=2.67^a$}
\tablecolumns{9}
\tablewidth{0pc}
\tablehead{
  \colhead{ID}  & 
  \colhead{$\Delta V^b$}  & 
  \colhead{FWHM$_{\text{CO}}$}   & 
  \colhead{$\mu I_{CO(3–2)}$} & 
  \colhead{H$_{\text{F160W}}^c$} & 
  \colhead{$\mu S_{\text{3 mm}}$} & 
  \colhead{$\mu S_{\text{6 GHz}}$} & 
  \colhead{$\mu L_{CO}^{\prime}$} & 
  \colhead{$\mu$}
\\[-2ex]
  \colhead{} & 
  \colhead{km s\(^{-1}\)} & 
  \colhead{km s\(^{-1}\)} & 
  \colhead{Jy km s\(^{-1}\)} & 
  \colhead{AB} & 
  \colhead{$\times10^{-4}$ Jy} & 
  \colhead{$\times10^{-4}$ Jy} & 
  \colhead{$10^{10}$ K km s\(^{-1}\)pc\(^2\)} & 
  \colhead{}
}
\startdata
1ab   & \(-255\pm 5\)  & \(417\pm 13\)  & \(13.03\pm 0.35\)            & \(22.29\pm 0.06\)  & \(8.68\pm 0.23\)  & \(3.29\pm 0.46\)  & \(47.0\pm 1.3\)  & \(14\pm5\)  \\ 
1c    & \(-230\pm 8\)  & \(447\pm 19\)  & \(4.47\pm 1.64\)                & \(23.68\pm 0.03\)  & \(3.35\pm 0.31\)  & \(1.58\pm 0.24\)  & \(16.1\pm 5.9\)  &  \(8.2\pm1\)  \\
2     & \(12\pm 30\)   & \(661\pm 79\)  & \(4.15\pm 0.97\)                & \(23.01\pm 0.05\)  & \(2.77\pm 0.16\)  & \(<0.08\)         & \(15.0\pm 3.5\)  &  \(4.2\pm0.1\)  \\
3     & \(-500\pm 11\) & \(434\pm 26\)  & \(2.06\pm 0.11\)                 & \(21.74\pm 0.02\)  & \(2.40\pm 0.45\)  & \(1.93\pm 0.45\)  & \(7.42\pm 0.40\) & \(3.2\pm0.1\)   \\ 
4a    & \(216\pm 14\)  & \(275\pm 67\)  & \(1.13\pm 0.19\)                & ...              & ...              & ...              & \(4.07\pm 0.69\) &  \(4.6\pm0.6\) \\
4b    & \(242\pm 16\)  & \(211\pm 35\)  & \(0.80\pm 0.11\)               & ...              & ...              & ...              & \(2.88\pm 0.40\) &  \(3.4\pm0.5\) \\
4c    & \(283\pm 14\)  & \(381\pm 38\)  & \(1.50\pm 0.13\)                 & ...              & \(0.15\pm 0.04\)  & ...              & \(5.41\pm 0.47\) &  \(3.7\pm0.1\) \\
5     & \(94\pm 15\)   & \(275\pm 34\)  & \(1.01\pm 0.11\)               & \(23.12\pm 0.03\)  & ...              & ...              & \(3.64\pm 0.40\) &  \(6.4\pm0.4\) \\
6.1ab & \(55\pm 4\)    & \(225\pm 9\)   & \(5.21\pm 0.19\)                & \(22.93\pm 0.02\)  & ...              & ...              & \(18.8\pm 0.7\)  &\(23\pm 12\)   \\
6.1c  & \(98\pm 9\)    & \(216\pm 20\)  & \(0.46\pm 0.08\)                 & ...              & ...              & ...              & \(1.66\pm 0.29\) &  \(2.5\pm0.1\) \\
6.2ab & \(66\pm 8\)    & \(255\pm 18\)  & \(2.00\pm 0.12\)                & ...              & ...              & ...              & \(7.21\pm 0.43\) &  \(21\pm7\)   \\
6.2c  & \(-57\pm 17\)  & \(253\pm 40\)  & \(0.08\pm 0.05\)               & ...              & ...              & ...              & \(0.29\pm 0.18\) &  \(2.7\pm0.1\)    \\
7\textsuperscript{d} & \(76\pm 14\)   & \(389\pm 33\)  & \(2.09\pm 0.16\)       & \(21.89\pm 0.02\)  & \(0.64\pm 0.07\)  & ...              & \(7.53\pm 0.58\) & \(3.1\pm0.3\)   \\
8.1   & \(-355\pm 21\) & \(457\pm 39\)  & \(1.75\pm 0.16\)       & \(21.44\pm 0.02\)  & \(1.12\pm 0.16\)  & \(<0.06\)         & \(6.31\pm 0.58\) & \(2.9\pm0.1\)    \\
8.2   & \(-255\pm 45\) & \(786\pm 105\) & \(0.83\pm 0.10\)           & \(23.16\pm 0.01\)  & \(0.54\pm 0.12\)  & \(0.19\pm 0.06\)  & \(2.99\pm 0.36\) & \(2.9\pm0.1\)  \\
9     & \(306\pm 23\)  & \(354\pm 54\)  & \(0.91\pm 0.12\)          & \(24.65\pm 0.05\)  & ...              & ...              & \(3.28\pm 0.43\) &  \(2.6\pm0.1\)   \\
10    & \(240\pm 35\)  & \(404\pm 81\)  & \(0.64\pm 0.11\)               & \(22.09\pm 0.01\)  & ...              & ...              & \(2.31\pm 0.40\) & \(2.3\pm0.1\)   \\
11    & \(-91\pm 22\)  & \(337\pm 58\)  & \(0.63\pm 0.06\)           & \(21.74\pm 0.02\)  & ...              & ...              & \(2.27\pm 0.22\) & \(1.5\pm0.1\)  \\
\enddata
\tablenotetext{a}{The 18 images are undetected in the Gemini photometry except for ID7 ($m_{\text{r$^{\prime}$}}= 24.55$ mag, $m_{\text{z$^{\prime}$}}=23.45$ mag) and ID8.1 ($m_{\text{z$^{\prime}$}}=24.13$ mag).}
\tablenotetext{b}{Velocity offsets relative to the mean redshift, $z = 2.666.$}
\tablenotetext{c}{Photometry not corrected for magnification, $(\mu)$.}
\tablenotetext{d}{This only includes the emission from 7.1, see ID7 description in Section \ref{Sec:The Image Systems}.}

\end{deluxetable*}

\begin{figure*}
	\centering\includegraphics[scale =0.615]{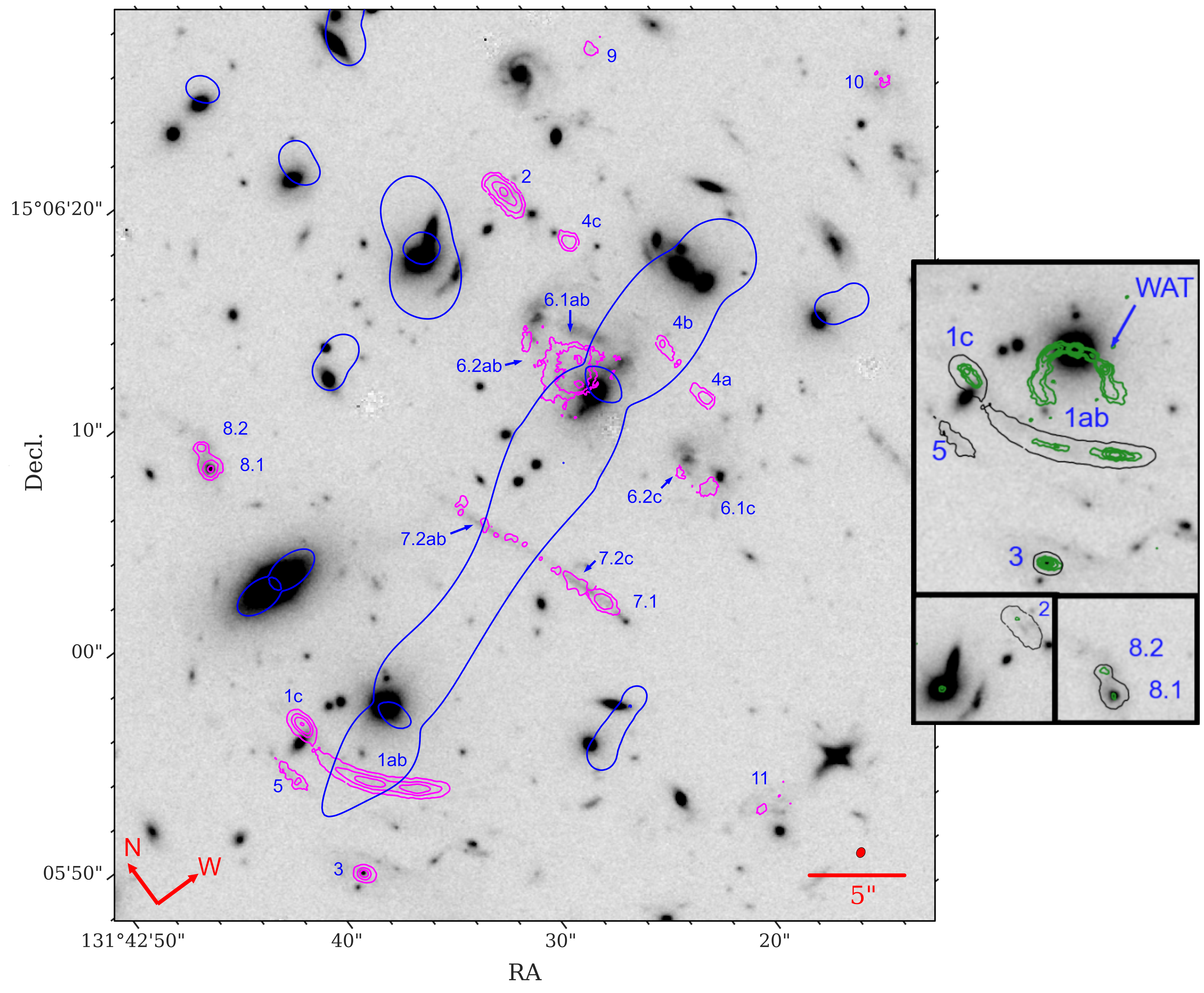}
	\caption{The J0846 field. Main panel: The 18 CO(3–2) moment-0 detections (magenta contours) at $z=2.660-2.669$ (J0846.BG) plotted over the HST/F160W imaging. The red ellipse in the bottom right represents the synthesized beam. The CO emission that makes up the ID7.2 images are only detected in $<3$ channels, but we display them here as they have crucial implication on the lens model (see ID7 description in Section \ref{Sec:The Image Systems}).  
    In the blue contour we provide the critical curve at $z=2.67$ of the lens model. Right panels: Insets across J0846 that have VLA 6 GHz continuum emission (green contours) and outline (black) of the moment-0 map. This includes two foreground cluster (J0846.FG) galaxies, one of which hosting a radio wide-angle tail.} 
	\label{fig:3}
\end{figure*}

\section{The Lens Model} \label{Sec:The Lens Model}
To construct the lens model we used \texttt{GLAFIC} \citep{Oguri2010, Oguri2021}, employing a parametric approach to predict the total baryonic and dark matter 2D mass distribution. The model is constrained by two primary inputs: the observed positions of the image systems (Section \ref{Sec:The Image Systems}) and the identified foreground cluster (Section \ref{Sec:The Foreground Cluster}). The model construction is essential to studying the intrinsic properties of these lensed CO detections. An initial preliminary lens model was constructed using the light traces mass (LTM) approach \citep{Zitrin2009, Zitrin2015}, which aided in image system identifications and provided insights into the cluster's lensing properties.

\subsection{The Image Systems} \label{Sec:The Image Systems} We identify 11 unique sources that are lensed into the 18 CO detections at $z=2.67$ observed with ALMA. Four of the 11 sources are lensed into multiple-imaged systems, while the remaining seven are only singly-imaged (weak-lensing). We denote counter-images of a system using an alphabetical sequence (i.e., a, b, c, d),  identified by corroborating counterimages via color, spectroscopic redshift, morphology, and model-predicted locations. There are three additional multiply-imaged systems without any spectroscopic redshift measurement, identified solely on F160W imaging. Their redshifts are left as free parameters to be inferred by the lens model geometry. The image systems are described below (see Figure \ref{fig:3}), starting with the multiply-imaged CO systems: 
\begin{figure*}
	\centering\includegraphics[scale =0.415]{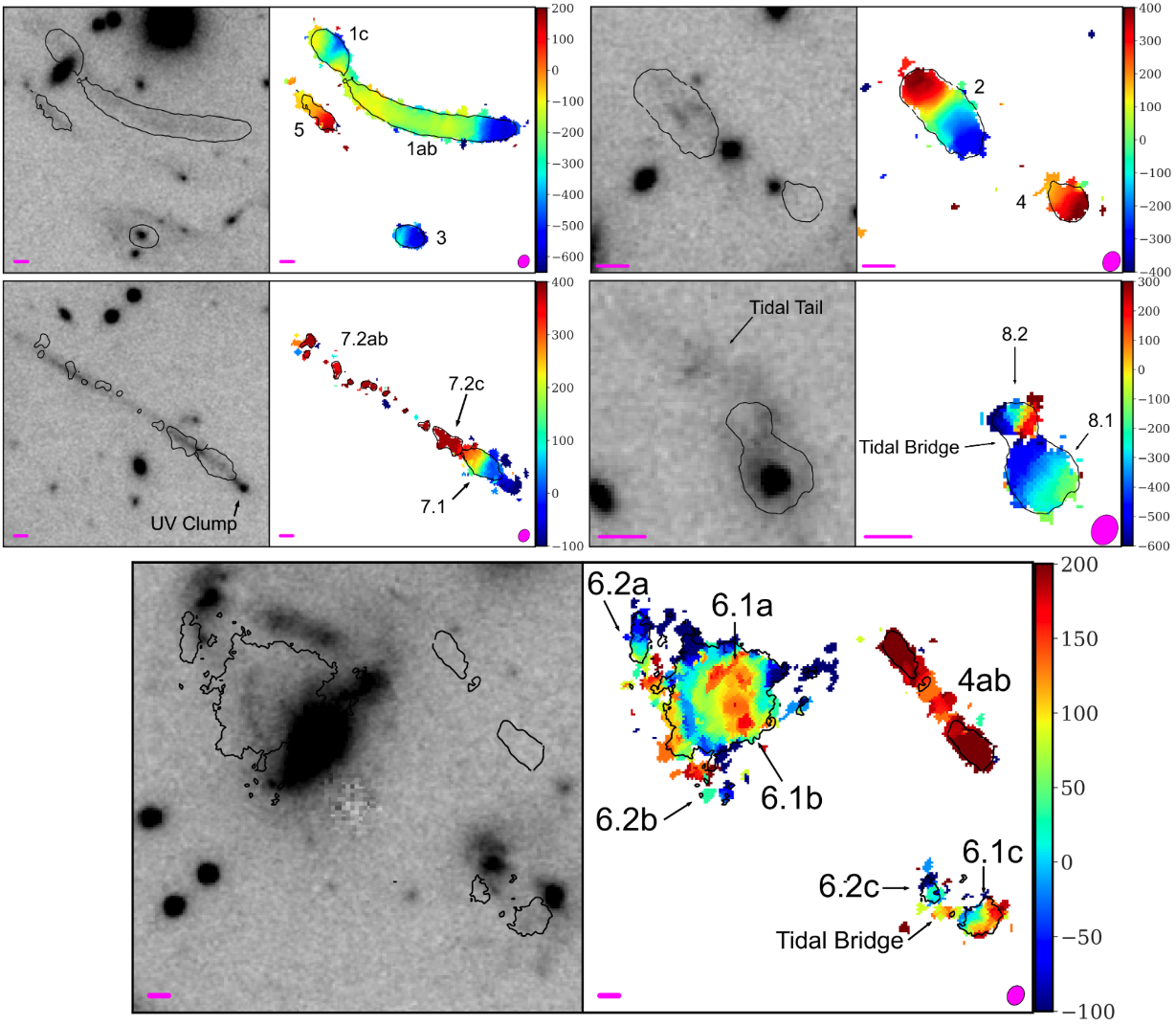}
	\caption{CO(3–2) moment-1 velocity maps for various sources in the J0846.BG. The left panel of each set, displays \textit{HST} F160W image with an 3$\sigma$ contour outline of the integrated CO(3–2) line moment-0 maps.
    On the right panels we show the moment-1 velocity maps with the same moment-0 contour overlaid. The color bar on the right of each set represents the corresponding velocities in km s$^{-1}$. In each stamp we provide a scale bar equating to $1^{\prime\prime}$ in the bottom left and the synthesized beam in the bottom right, with the same orientation as in Figure \ref{fig:3}.
    }
	\label{fig:4}
\end{figure*}

\underline{\textbf{ID1:}} 
This system consists of three CO(3–2) images, as the source crosses a fold caustic producing a pair of merging images in the form of a long arc (1ab), as well as a third image (1c). It is also faintly detected in the F160W imaging (see Table \ref{Table:1}). 

\par
\underline{\textbf{ID4:}} 
This system consists of three CO(3–2) images, as the source is likely nearby a fold caustic producing a pair of merging images on the critical curve (4ab) as well as an expected third image (4c). The symmetry and parity flip of the images of the 4ab image pair provides a powerful local constraint on the critical curve's position.  
\par
\underline{\textbf{ID6:}}
In the moment-0, CO(3–2) line integrated map, image ID6.1ab appears as an extended source detection with approximate angular diameter of $\theta \sim $ 2.5$^{\prime\prime}$. Further analysis shows that the detection consists of two spatially distinct emission peaks, 6.1a and 6.1b. The CO moment-1 velocity map exhibits a symmetric structure along the axis bisecting the two peaks, crossing the critical curve (see Figure \ref{fig:4}). Nearby 6.1ab, a companion structure is detected (6.2ab), a thin arc, $<1^{\prime\prime}$ away, that similarly merges across the critical curve. We locate the expected third counterimages of both components as ID6.1c and ID6.2c which both follow a consistent kinematic profile. We refer to ID6.1/ID6.2 as a single system (ID6) despite what may be multiple kinematic components in the line profile.
\par

\par
\underline{\textbf{ID7:}} In the F160W imaging ID7 appears as a compact bright source with an extended tail that gets lensed into a straight arc, producing a fold image. The peak of the CO emission is offset by $<1.8^{\prime\prime}$ to this bright stellar component. In the CO data cube, we do detect this straight arc coinciding with the optical/IR equivalent, but only across a few channels or $<150$ km s$^{-1}$, towards the redder end of the line profile (see Figure \ref{fig:4}). The interpretation is that most of the galaxy resides just outside of the caustic only resulting in a single image (ID7.1), while a small region of the source crosses the caustic eventually producing multiples images manifesting into a straight arc (ID7.2a, ID7.2b). 

\par
\begin{figure}[h]
\centering\includegraphics[scale =0.125]{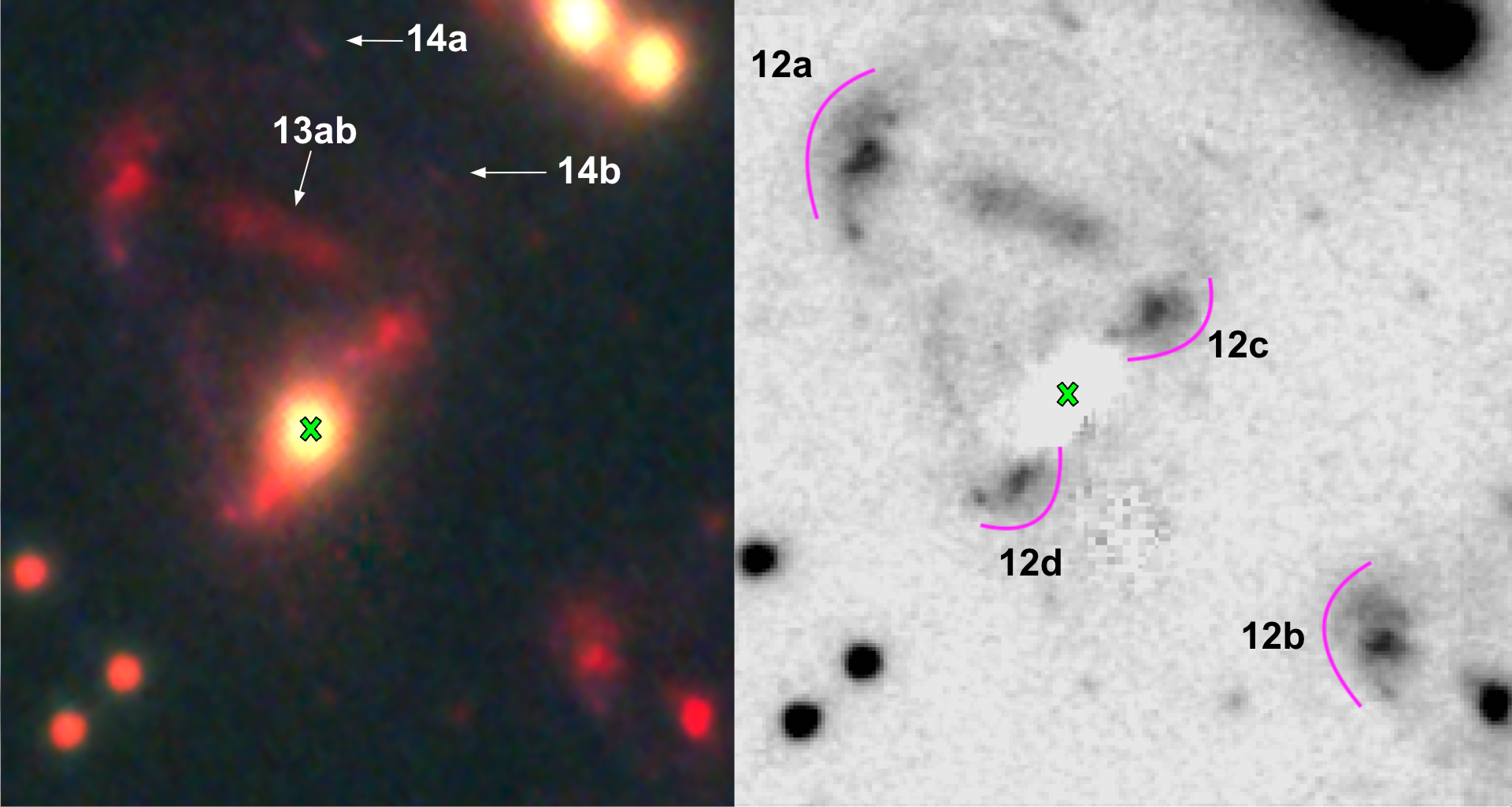}
\caption{ The identified non-CO emitting image systems (ID12, ID13, ID14). Left Panel: color image using r$^{\prime}$, z$^{\prime}$, F160W. Right Panel: F160W image with the foreground galaxy (green `$\times$' symbol) subtracted utilizing \texttt{GALFIT}. Labeled are the four images of the Einstein Cross, illustrating the flip in parity with the comma-like morphology. For reference, the FOV corresponds to the bottom panel of Figure \ref{fig:4}, in the vicinity of image systems 4 and 6.} 
 
\label{fig:5}
\end{figure}

\underline{\textbf{Non-CO Image Systems:}} 
        There are three multiply-imaged systems that are identified in the \textit{HST}/Gemini imaging that lack counterpart ALMA detections. All three of these image systems are located in the northwest region of the cluster lens, within $\sim 5^{\prime\prime}$ of 6ab. We display their image designations in Figure \ref{fig:5}, which corresponds to the FOV of the bottom panel of Figure \ref{fig:4} for reference. These image systems do not have spectroscopic redshifts and are therefore left as free parameters in the model.
\par
\underline{\textbf{ID12, The Einstein Cross:}}
 We identify a rare strong lensing event producing a quadruply-imaged system consisting of two relatively isolated, well-separated images (12a, 12b) and two images which are close in projection to a cluster elliptical galaxy (12c, 12d). The foreground galaxy was subtracted off by modeling its luminosity with a Sérsic profile \citep{Sersic_1963} utilizing the \texttt{GALFIT} \citep{Peng_2002} software. All four images have a similar red color (albeit, only detected in F160W while undetected in r$^{\prime}$ and z$^{\prime}$) and a distinct morphology consisting of a compact core with a spiral-shaped arm extending outward like a `comma' (Figure \ref{fig:5}). 
Its resolved structure also illustrates the expected image parity flip across the critical curve, where the `comma' shape flips side (i.e., 12a and 12b are mirror images of 12c and 12d). Alternatively, such image formation could be due to the source lying near a hyperbolic umbilic \citep[e.g.,][]{Meena_2023}. 

\par
\underline{\textbf{ID13 and ID14:}} 
Both the ID13 and ID14 systems are identified with a pair of counterimages. They appear to be examples of fold images, as evidenced by both pairs exhibit merging images. All images are only detected in F160W while completely dropping out in the bluer r$^{\prime}$ and z$^{\prime}$ imaging.
\par
\underline{\textbf{Single-Image CO Systems:}} 
The remaining seven of 11 CO(3–2) systems are only singly-imaged (ID: 2, 3, 5, 8, 9, 10, 11), but nonetheless still provide valuable constraints on the foreground mass distribution. The information can inform the shape of the caustic providing an upper limit on the locally-derived lensing potential. Given a single image designation, a penalty term is enforced into the likelihood function to discourage the model from producing multiple images for the source. For example, ID8.1/ID8.2 is located to the NE of the bright elliptical perturber at $z=0.357$, approximately $\sim 7^{\prime\prime}$ away. In this case, it provides insight into the lensing potential induced by the nearby perturber (see Section \ref{Sec:The Perturber: A Bright Elliptical}). Similarly, as discussed, the majority of the CO(3–2) emission from ID7 (ID7.1) is not multiply imaged, while only a small portion (ID7.2) gets lensed into the straight arc. This information provides a powerful constraint to the shape of the caustic curve, pinpointing its position nearby the source.

\subsection{The Foreground Cluster} \label{Sec:The Foreground Cluster}

\begin{figure}[h]
\centering\includegraphics[scale =0.207]{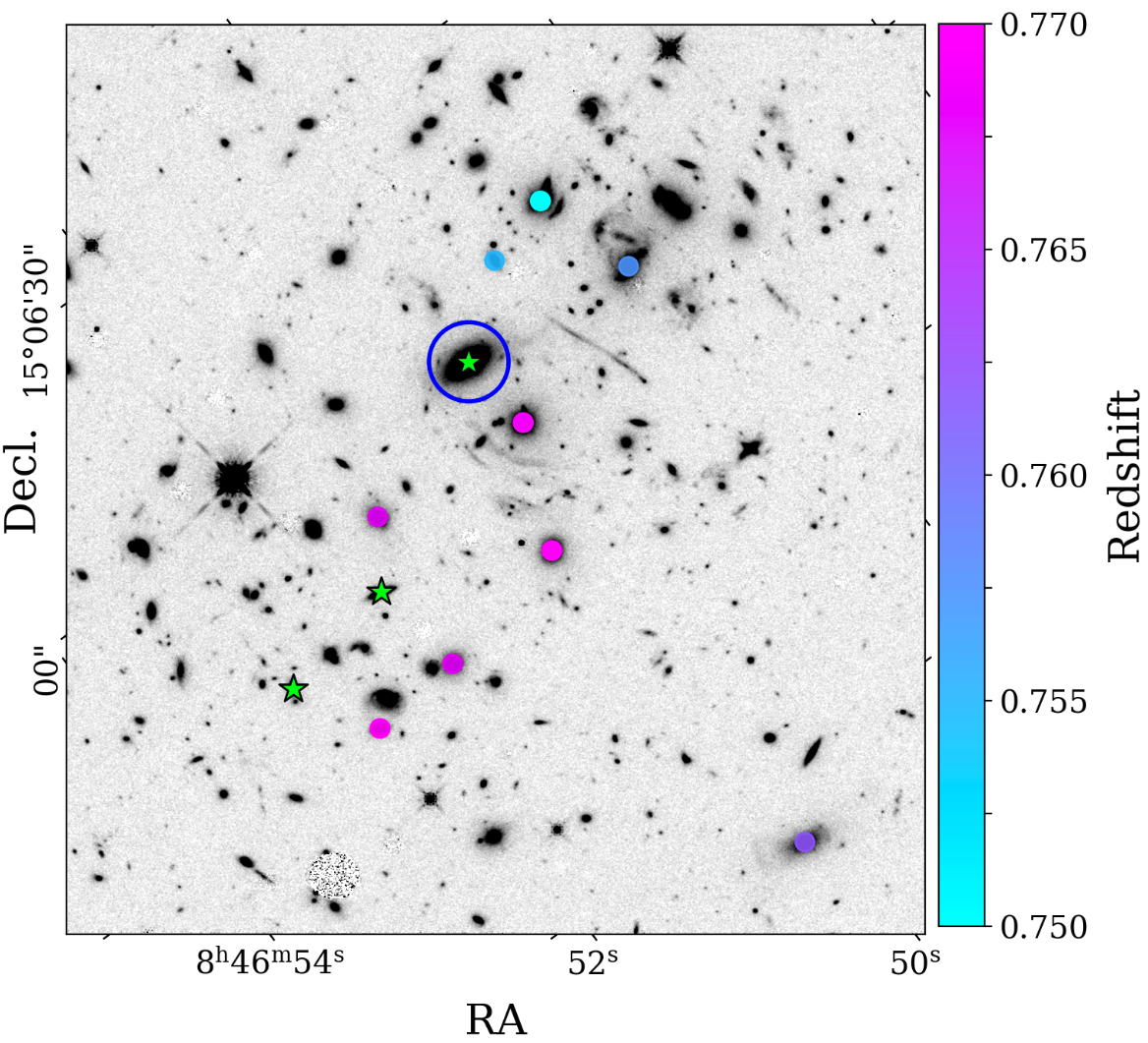}
\caption{Optical spectroscopic redshifts plotted over the F160W image. The cluster (J0846.FG) member galaxies within $z=0.75-0.772$ are shown with their corresponding color-scale (filled circles). The green stars indicate objects that have redshift measurements outside the cluster redshift, all at $z<0.4.$ The bright elliptical circled in blue indicates a foreground perturber at $z=0.357$.}
\label{fig:6}
\end{figure}

\begin{figure}[h]
\centering\includegraphics[scale =0.31]{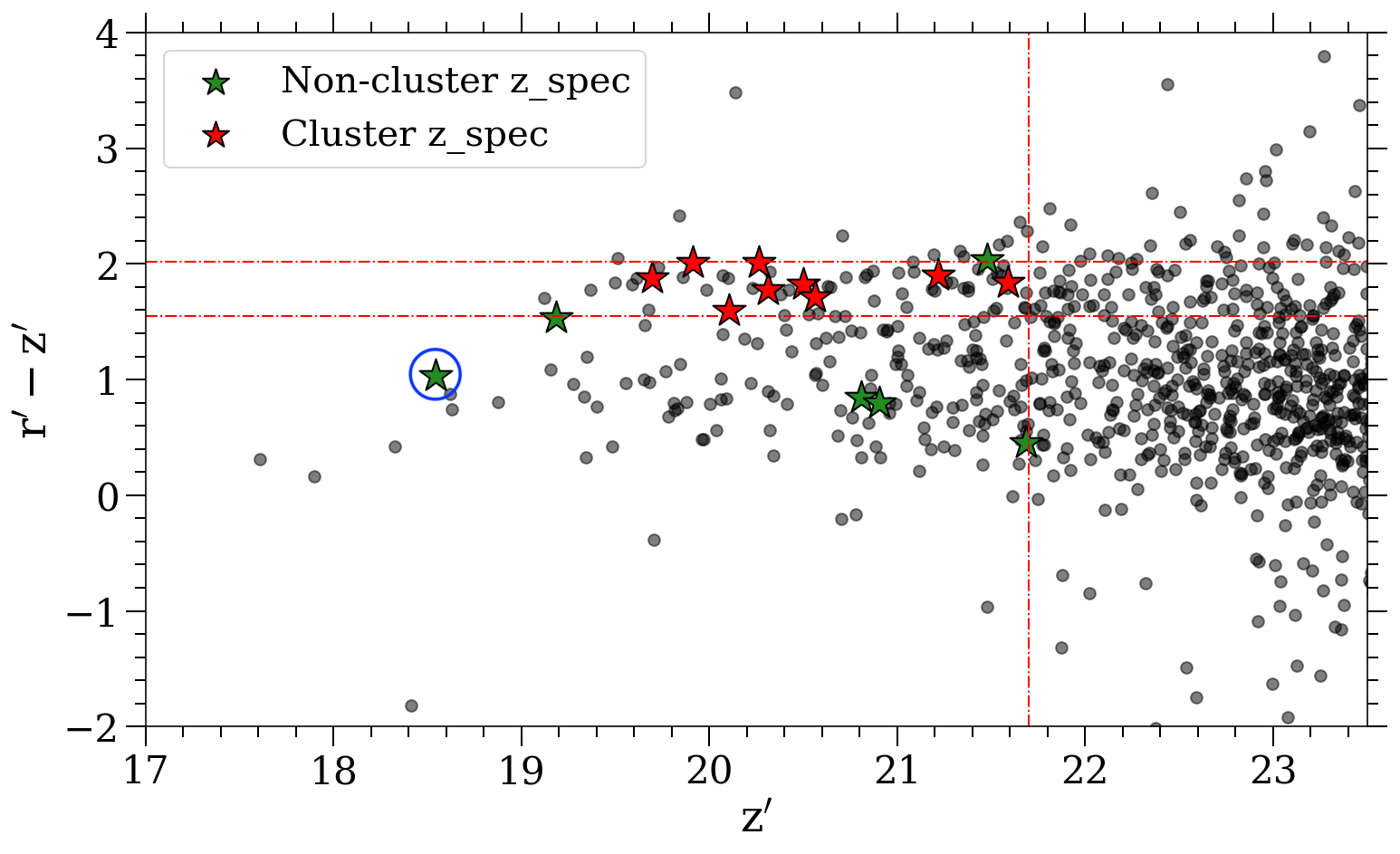}
\caption{Color-magnitude diagram using r$^{\prime}$ and z$^{\prime}$, covering rest-frame 3500$\text{\AA}$ and 5068$\text{\AA}$ at $z=0.77$, respectively, effectively bracketing the 4000$\text{\AA}$ break. All stars have also been removed on the plot via FWHM selection criteria. The red and green stars are spectroscopic objects at the cluster redshift and outside of it, respectively. The spec$-z$ object circled in blue indicates the bright foreground elliptical at $z=0.357$. The red horizontal lines indicate our color cut selection to identify the cluster red sequence also imposing a magnitude cut at $m_{\text{z$^{\prime}$}}=21.7$. }
\label{fig:7}
\end{figure}

With the available optical spectroscopy, we identify nine cluster member galaxies in J0846.FG within the redshift range of $z=0.750-0.772$ (see Figure \ref{fig:6}), or $z=0.766\pm0.002$ utilizing the biweight estimator \citep{Beers1990, Ruel_2014}. The spectroscopic redshift of one cluster member was contributed from our Gemini observations, six were measured from the MUSE data, and six were obtained from the FORS2 data with several members having multiple, confirming measurements (Table \ref{Table:A1}). Given the incomplete spectroscopic coverage of the field (see Appendix \ref{Appendix:Optical Spectroscopy}), we extrapolate the full membership by identifying the cluster's red sequence via a r$^{\prime}$ and z$^{\prime}$ color-magnitude diagram method \citep{Gladders_2000}, guided by the confirmed spectroscopic members (Figure \ref{fig:7}).

\par

Although one may perceive of a redshifted and blueshifted component in J0846.FG (see Figure \ref{fig:6}), existing spectroscopic data is insufficient to statistically claim a bimodal velocity distribution. Nevertheless, the scope of this paper is to provide the first characterization of the foreground cluster to construct a lens model, where a bimodal mass distribution (i.e., a merging cluster) will have direct implications on its lensing properties. The emergence of a straight arc, ID7.2a/b (see Section \ref{Sec:The Image Systems}), observed in-between the SE and NW regions of the cluster further supports a merger scenario. This strong lensing feature is typically produced from a caustic configuration generated by two merging mass components, e.g., ``La Flaca'' in the El Gordo cluster \citep{Diego_2023, Frye_2023}. Additionally, the discovery of a radio wide-angle tail (WAT, see right panel of Figure \ref{fig:3}) hosted by one of the confirmed cluster galaxies is often linked to non-relaxed systems (see Appendix \ref{Sec:A Radio Wide Angle Tail Galaxy} for further discussion).

\begin{figure*}
\centering\includegraphics[scale =0.265]{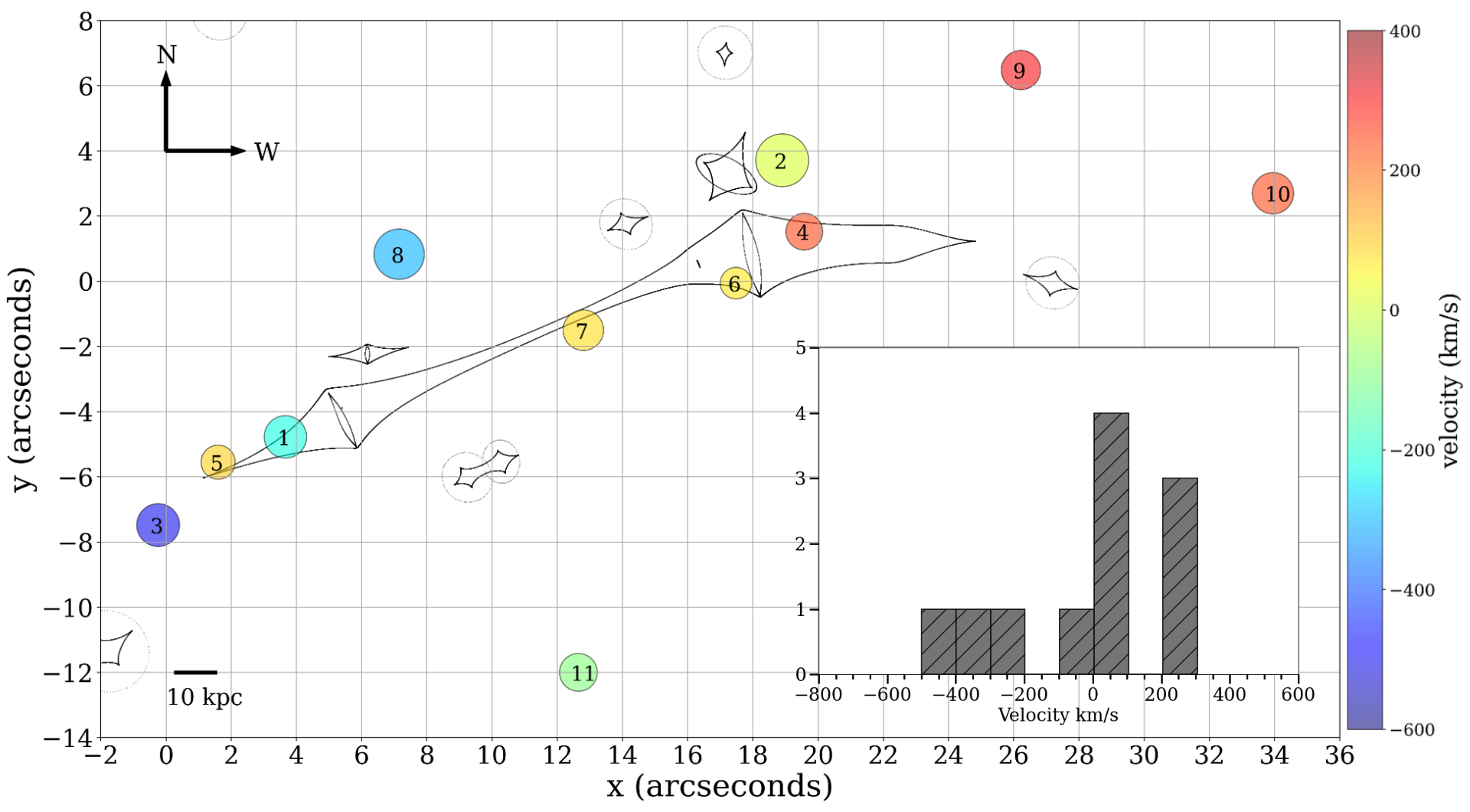}
\caption{ 
The source plane reconstruction of J0846.BG, 11 CO(3–2) member galaxies at $z=2.67$ plotted over the corresponding caustic curve (in black) from the lens model. Their respective color indicates the velocity offset relative to the average velocity for all member galaxies. The measured FWHM from each source's CO profile are reflected in their relative sizes on the plot. In the bottom right inset we display the velocity histogram of all galaxies. We find that these 11 CO galaxies resides in a physical area of 150 kpc $\times$ 280 kpc all within a velocity offset of \(\Delta V \approx 800\) km s\(^{-1}\), and velocity dispersion of \(\sigma_v = 246 \pm 72\) km s\(^{-1}\). The center of the plane at $x=0$, $y=0$ corresponds to the WCS coordinate at RA=131.70877, Decl=15.09936.}
\label{fig:8}
\end{figure*}

\subsection{Parametric Mass Profiles}

To represent the cluster members of J0846.FG in the lens model we utilize pseudo-Jaffe ellipsoid profiles \citep{Jaffe_1983, Keeton_2002}, scaled by their relative observed luminosities measured in the F160W photometry. The density profile is a double power-law behaving $\rho \propto r^{-2}$ in the inner region and $\rho \propto r^{-4}$ as $r \to \infty$. Unlike the original Jaffe profile where $\rho \propto r^{-2}$ as $r \to 0$, the pseudo-Jaffe profile introduces a core radius ensuring that the density is finite at the center. For 5/33 cluster members, we instead adopt an isothermal ellipsoid density following a $\rho \propto r^{-2}$ profile, which similarly implements a core radius. Here we allow their velocity dispersion, ellipticity and position angle to be independently fit, resulting in a more robust density profile model. These cluster members are near in projection to key image systems, for example, the giant arc (ID1) and Einstein cross (ID12) where the increased flexibility is necessary. To account for the cluster dark matter halo we introduce a Navarro-Frank-White (NFW) profile \citep{Navarro_1996}, centered on the proposed BCG, hosting the radio WAT, allowing its position to have some freedom during the model fitting. \texttt{GLAFIC} is equipped to solve the lens equation for multi-plane configurations, enabling us to explicitly model the foreground perturber that is situated near the center of the cluster (see Figure \ref{fig:4}). For this bright elliptical we place another NFW profile at $z=0.357$, utilizing the estimate of its dynamical mass based on the line width measurements (see Section \ref{Sec:The Perturber: A Bright Elliptical}) as a prior for fitting in the model.


\subsection{The J0846 Model Results}
We use the Python package \texttt{emcee} \citep{Foreman-Mackey_2013} to implement Markov chain Monte Carlo (MCMC) approach to sample the posterior distribution of the large parameter space with thousands of chains \citep[see][]{Liu_2023}. We construct the likelihood function based on the ability of the model to reproduce the positions of the observed multiply imaged systems. Singly-imaged galaxies (i.e., ID2, ID5, ID8) are also used as constraints, as we enforce an additional penalty if the model predicts multiple images. The best fit model reproduces the angular positions of identified multiply imaged systems to an rms separation of $0.14^{\prime\prime}$. For the CO images, the magnification factors are calculated as the average value on the magnification map within each aperture used for the line extraction method in Section \ref{Sec:The CO Detections at $z=2.67$}. For non-CO images we measure the magnification factor corresponding to their coordinate designation. Magnification factors are provided in Table \ref{Table:2}, as well as geometric-predicted redshifts for ID12, ID13, ID14 in Table \ref{Table:A2} of the appendix.  
To determine the statistical uncertainties on the magnification factors and geometric redshifts we randomly resampled 1000 model iteration from the MCMC posterior distribution. We acknowledge that this does not account for the systematics that arise from cluster lens modeling, that likely dominate the overall uncertainties. A handful of studies have explored the dependence of different lens models for select clusters \citep[e.g.,][]{Johnson_2016, Meneghetti_2017, Pascale_2024, Liu_2025}, suggesting systematic errors on magnification, redshift, and foreground mass are primary driven by insufficient number of multiply imaged system used as constraints; especially those lacking spectroscopic redshifts. For example, \cite{Johnson_2016} showed by constructing a variety of models for a simulated cluster that the accuracy of magnifications increases with the number of image systems, where $>$ 20 consistently yields errors of 2\%. In the case of J0846, although several spectroscopic multiply imaged systems are identified they are all at a very similar redshift, and may not be able to break the mass-sheet degeneracy \citep[see][]{Schneider_1995}. A full analysis exploring the full systematics of the lens model will be left to future work as additional constraints will be obtained overtime. 
\par
With the constructed lens model we de-lens the 18 CO images back into the source plane. Here we report that J0846.BG is made up of 11 lensed DSFGs all within a velocity range of $\Delta V\approx800$ km s$^{-1}$ ($z=2.660-2.669$) and a projected source plane extent of 280 $\times$ 150 kpc. We show their positions in the source plane relative to the lensing caustic as well as illustrating their relative velocity offsets and measured FWHM$_{\text{CO}}$ in Figure \ref{fig:8}.  \\

\section{Properties of a Protocluster Core Candidate} \label{Sec:Global Properties of a Protocluster Candidate}
%
We discuss the discovery of what may be one of the most highly star-forming, gas- and dust-rich protocluster core candidates, J0846.BG. This system is uniquely situated along our line-of-sight, which enables us to have the rare advantage of examining a DSFG-rich protocluster candidate being strongly magnified. Here we outline the initial estimates of the sum of the global galaxy properties and prominent features observed among the members of J0846.BG.



\subsection{Star Formation Activity} \label{sec:Star Formation Activity}
The global SFR in J0846.BG has been determined from prior calculations of the integrated IR-luminosity (8-1000$\mu$m) \citep{Harrington2021, Berman2022} using the available photometry -- which is limited by \textit{Planck} measurements with a PSF of \(\sim 2^{\prime}\). The SED fits to the far-IR-to-mm dust continuum measurements in \citet{Berman2022} use a representative SED template \citep[see e.g.,][]{Efstathiou_2000, Harrington2016}, with photometry measurements of only one system, resulting in an apparent \(\mu\mathrm{SFR} \sim 14{,}000~\mathrm{M}_{\odot}~\mathrm{yr}^{-1}\). This likely underestimates the global SFR for this system that is composed of at least 11 dusty sources, therefore a single galaxy template fit is likely inappropriate and is considered a lower limit. \citet{Harrington2021} fit the dust photometry and CO using the sum of the CO line emission from pointed single dish observations, still complicating the analyses of multiple objects. Therefore we opt to fit a modified blackbody to the observed dust SED using the sum of all unresolved dust continuum data points \citep[from][]{Harrington2021, Berman2022}. We find a physical source area of $A=482^{+331}_{-291}$ kpc$^{2}$, a dust temperature of $T_d=47.1^{+8.8}_{-6.4}$ K, a dust mass of $M_d=2.21^{+0.62}_{-0.53}\times10^{10}\textup{ M}_\odot$, and an emissivity index of $\beta=1.63^{+0.19}_{-0.14}$ (See Appendix \ref{Appendix:Global Modified Blackbody Dust SED} for further details). We determine the total apparent IR luminosity by integrating 
the SED fit from $8-1000\mu$m, finding a value of \(\mu\mathrm{L}_{\mathrm{IR}} = 3.75^{+2.16}_{-1.21} \times 10^{14}~\textup{L}_{\odot}\) and a \(\mu\mathrm{SFR}=39900^{+23000}_{-12900}~\mathrm{M}_{\odot}~\mathrm{yr}^{-1}\), reflecting the large systematic uncertainties in determining the global SFR\footnote{A Kroupa corrected relation for IR luminosity to SFR is applied \(\mathrm{SFR} = \mathrm{L}_{\mathrm{IR}} / (9.4 \times 10^{9}~\textup{L}_{\odot})\) \citep{Kennicutt_1998}} for this field. 
We acknowledge that a single SED fit may still be insufficient in characterizing the dust emission across multiple sources with potentially varying ISM properties, and thus spatially resolved follow-up will better constrain the SFR for each individual member.

\par
\begin{figure}[h]
\centering\includegraphics[scale =0.325]{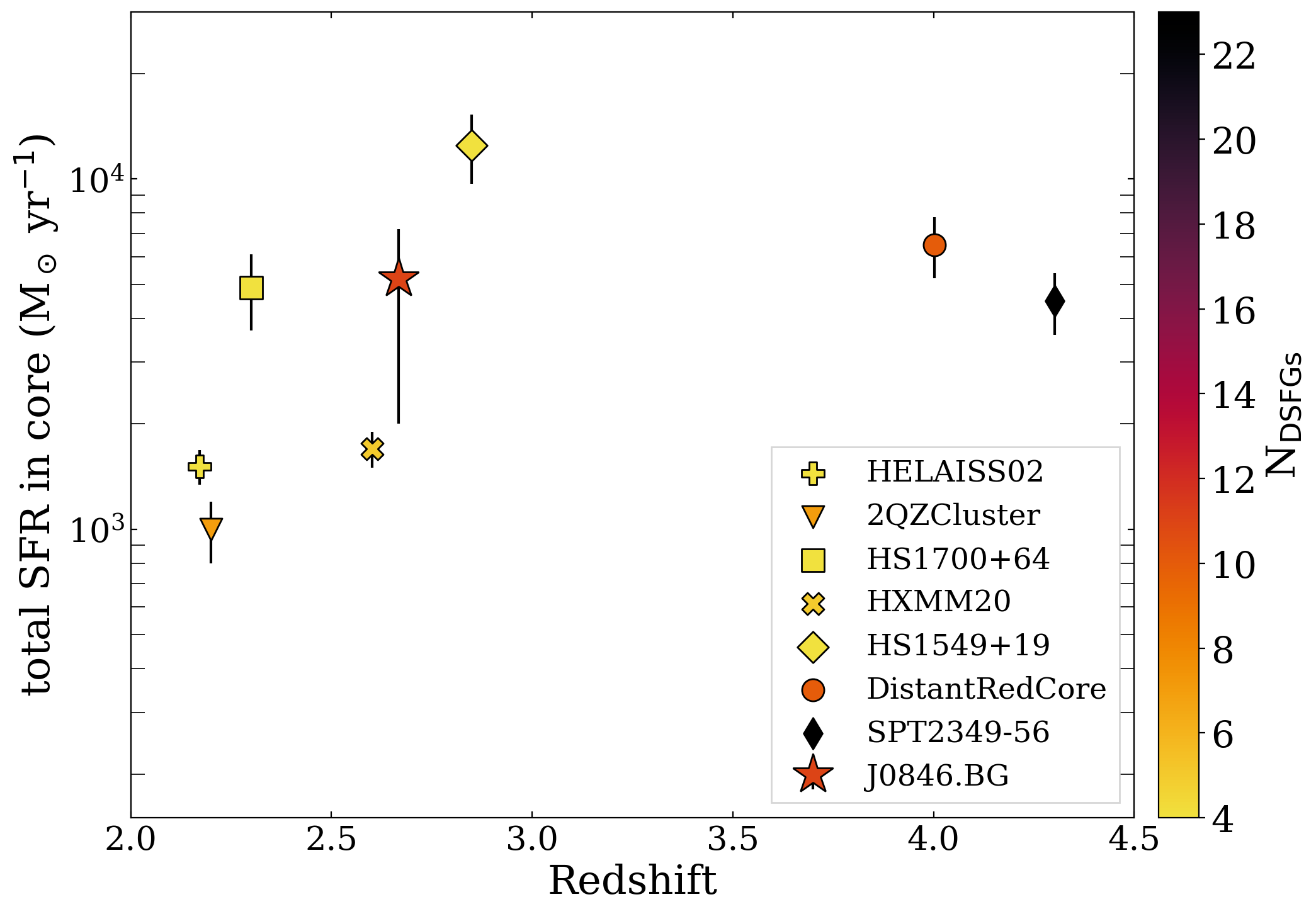}
\caption{ 
For comparison with J0846.BG (plotted with a star), the total SFRs (y-axis) are shown for different protocluster cores with known DSFGs at varying redshifts (x-axis) reported in \citet{alberts_2022}. The color-bar denotes the number of DSFGs that are identified for each system.} 
\label{fig:9}
\end{figure}

From the magnification map generated with the lens model, we estimate a mean magnification of $\sim2$ over the entire region, suggesting an intrinsic SFR of $\sim20,000~\textup{M}_{\odot}$ yr$^{-1}$. This exceedingly high value is biased by differential magnification effects and does not properly account for the relative contribution of each source to the total SFR. Therefore we consider the magnification based on the ratio of the total intrinsic (Table \ref{Table:3}) versus observed CO line luminosities (Table \ref{Table:2}), of $\mu=7.7\pm1.5$, resulting in a total intrinsic value up to SFR $ = 5200^{+3200}_{-2000}~\textup{ M}_{\odot}$ yr$^{-1}$. For comparison, we plot the known protocluster core systems with DSFGs between $z=2-5$ in Figure \ref{fig:9} \citep[][and references therein]{alberts_2022}. For example, SPT2349-56 is one of the most DSFG-rich protoclusters ever discovered, consisting of 23 galaxies at $z = 4.304$ detected via CO and/or [CII] with a velocity dispersion of $\sigma_{\text{vel}}\sim410 $ km s$^{-1}$, all residing within $\sim300$ kpc, and a total SFR of up to $\sim4,480~\textup{M}_{\odot}$ yr$^{-1}$ in the core \citep{Miller_2018, Hill_2020}. This makes J0846.BG among the most starbursting protocluster core candidates to be reported \citep{alberts_2022}.

\subsection{Estimates of molecular gas and dust masses}\label{Sec:First order estimates of molecular gas and dust masses}
Here we estimate a fiducial molecular gas mass for each member from the derived CO(3–2) line luminosity, adopting a conservative CO-H$_2$ conversion factor. We adopt the lower value found in \cite{Harrington2021} of $\alpha_{\text{CO(3–2)}}\sim1\textup{ M}_{\odot} \ (\text{K km s}^{-1} \text{ pc}^2)^{-1}$. We then directly translate the measured, de-magnified $\text{L}_{\text{CO(3-2)}}^{\prime}$ to a total molecular ISM gas mass estimate for each system using this value. We acknowledge the significant uncertainties that arise from assuming such a value. \cite{Harrington2021} found, via a non-LTE radiative transfer analyses of low-to-high-J CO lines, that the conversion factor for the low-J CO lines, such as $\alpha_{\text{CO(1-0)}}$, $\alpha_{\text{CO(2-1)}}$, or $\alpha_{\text{CO(3-2)}}$ spans an order of magnitude for the PASSAGES sample of DSFGs, i.e. the parent sample for J0846.BG. Furthermore, the members of J0846.BG likely exhibit varied gas excitations, reflecting the diversity of their properties (Section \ref{Sec:A Diverse Population of Galaxies}). The value of $\alpha_{\text{CO(3–2)}}\sim1\textup{ M}_{\odot} \ (\text{K km s}^{-1} \text{ pc}^2)^{-1}$ thereby sets a lower limit for the molecular gas masses (Table \ref{Table:3}). The sum of these individual members yields a total molecular gas mass reservoir of M$_{ISM}= (2.0\pm0.3)\times10^{11}\textup{ M}_\odot$\footnote{ Corrected by the Helium abundance, a factor 1.36 \citep{Allen_1973}.}, within an order of magnitude of other reported protocluster cores \citep[e.g.,][]{Miller_2018, Oteo_2018,Champagne_2021}. Nonetheless, this value is likely an underestimate, where in future work we will conduct a resolved multi-J CO/[CI] line analysis to constrain the molecular gas excitation properties of the ISM for each member.



\par
We also estimate a fiducial dust mass for each source using the 3 mm photometry assuming a modified blackbody model. Although some DSFGs at $z=2-3$ have been shown to have dust opacities greater than unity out to 200 $\mu$m \citep{Harrington_2019}, to first approximations we assume all systems are in the optically thin regime\footnote{Further resolved dust SED analyses may uncover some systems similar to Arp220, with optically thick dust out to 1 mm (e.g., \cite{Scoville_2016})} at rest-frame $\sim1$ mm. Following, e.g., \cite{Hildebrand_1983}, the dust mass can be calculated using 

\begin{equation}
M_{\text{dust}} = \frac{  S_{\nu_{\text{obs}}}     D_{L}^{2}   }{  (1+z)  \kappa_{\nu_{\text{rest}}}  B_{\nu_{\text{rest}}}(T_{\text{dust}}) }
\end{equation}

\noindent where $S_{\nu_{\text{obs}}}$ is the 3 mm continuum flux, $\kappa_{\nu_{\text{rest}}} $ is the mass absorption coefficient and $B_{\nu_{\text{rest}}}$ is the Planck function for a given dust temperature, $T_{\text{dust}}$.

Without resolved multi-band dust SED measurements we adopt a fiducial value for the dust temperature of $T_{\text{dust}}= 44.7$ K (with a 20\% error) from \cite{Harrington2021}, the mean value derived from the larger PASSAGES parent sample (of which included J0846.BG). Dust temperatures in this sample ranged from $T_{\text{dust}}= 28-66$ K. This is greater than the typical value of $T_{\text{dust}}=25$ K for star-forming galaxies that is assumed to carry the bulk of the colder dust mass \citep{Scoville_2014}, which may be unsuitable for these dusty, starbursting objects. Furthermore, the mass absorption coefficient, $\kappa_{\nu_{\text{rest}}} $, may lead to an additional systematic uncertainty of up to a factor of two depending on assumptions made and choice of $\kappa_{\nu} $ \citep{Harrington2021}. Here we adopt the same assumptions as \citet{Harrington2021} following \citet{Draine_Lee_1984}, and fix the dust emissivity index, $\beta=1.8$. 
 Dust masses of each detected member are reported in Table \ref{Table:3}. We find molecular gas-to-dust mass ratios ranging $\sim90-700$, which may already indicate variation in $T_{\text{dust}}$ and $\alpha_{\text{CO}}$ among the sample. A spatially resolved dust and line SED analyses will further constrain these parameters.







\par
\subsection{Dynamical Mass Estimate} \label{Sec:Dynamical Mass Estimate}
We estimate the dynamical mass of the core structure using the measured velocity dispersion and the physical extent of the 11 galaxies with spectroscopic redshifts. First, to determine the mean redshift of these members, we utilize the biweight estimator \citep{Beers1990, Ruel_2014}, calculating a value of $z=2.6655\pm0.0008$. The velocity dispersion is calculated based on their proper velocities using the biweight sample variance, yielding $\sigma_{v}=246\pm72$ km s$^{-1}$. 
To estimate the dynamical mass of this core structure, we employ the virial theorem assuming a spherical mass distribution: 
\begin{equation} \label{Eq:Dyn Mass}
M_{dyn} = 3 \frac{R\sigma^2}{G}
\end{equation}
where R, is the maximum physical separation over all members, measured to be $280$ kpc. Here we estimate the dynamical mass to be M$_{dyn}=1.2\pm0.6\times10^{13} \textup{ M}_\odot$, similar in comparison to other reported protocluster cores \citep{Wang_2016, Miller_2018, Champagne_2021}, and likely a progenitor of a $\gtrsim2\times10^{14} \textup{ M}_\odot$ cluster at $z=0$ \citep{Chiang_2013}. This estimate does not account for the large systematics given it is unlikely that all 11 members are gravitationally bound to a single spherical halo that has fully virialized. We also do not consider for any peculiar motion contributing to the measured velocity, which may be significant for infalling members.  Furthermore, it is likely that the full extent of the J0846.BG has not yet been accounted for, where protoclusters are expected to span tens of cMpc scales \citep[e.g.,][see Section \ref{Sec:Challenges in Defining J0846 as a protocluster core}]{Muldrew_2015}, leading to an underestimate of the system's total dynamical mass.

\begin{deluxetable*}{cccccccc}
\label{Table:3}
\tablecaption{Intrinsic Properties of J0846.BG Members}
\tablecolumns{8}
\tablewidth{0pc}

\tablehead{
  \colhead{ID}  & 
  \colhead{x}  & 
  \colhead{y}   & 
  \colhead{$z$} & 
  \colhead{$L_{CO}^{\prime}$} & 
  \colhead{M$_{\text{gas}}$} & 
  \colhead{M$_{\text{dust}}$} & 
  \colhead{Type} 
\\[-2ex]
  \colhead{} & 
  \colhead{arcsec} & 
  \colhead{arcsec} & 
  \colhead{} & 
  \colhead{$10^{10}$ K km s\(^{-1}\) pc\(^2\)} & 
  \colhead{$10^{10}$\,\textup{M}$_{\odot}$} & 
  \colhead{$10^{8}$\,\textup{M}$_{\odot}$} & 
  \colhead{} 
}

\startdata
1\(^a\)    & 3.7 & \(-4.8\) & 2.663 & \(2.0 \pm 0.8\) & \(2.0 \pm 0.8\) & \(2.1 \pm 0.6\) & U \\ 
2         & 18.9 & 3.7 & 2.665 &  \(3.6 \pm 0.8\) & \(3.6 \pm 0.8\)  & \(3.4 \pm 0.8\) & D \\ 
3         & \(-0.2\) & \(-7.5\) & 2.660 &  \(2.3 \pm 0.2\) &  \(2.3 \pm 0.2\) & \(4.2 \pm 0.3\) & D \\ 
4\(^a\)    & 19.6 & 1.5 & 2.669 & \(1.5 \pm 0.1\) &  \(1.5 \pm 0.1\) & \(0.21 \pm 0.08\) & D \\ 
5         & 1.6 & \(-5.6\) & 2.666 & \(0.57 \pm 0.07\) &  \(0.57 \pm 0.07\) & ... & D \\ 
6\(^a\)    & 17.5 & \(-0.1\) & 2.666 & \(0.77 \pm 0.18\) & \(0.77 \pm 0.18\) & ... & M \\ 
7         & 12.8 & \(-1.5\) & 2.666 & \(2.4 \pm 0.3\) & \(2.4 \pm 0.3\) & \(1.1 \pm 0.4\) & D \\ 
8         & 7.2 & 0.8 & 2.662 & \(3.2 \pm 0.3\) & \(3.2 \pm 0.3\) & \(2.9 \pm 0.9\) & M \\ 
9         & 26.2 & 6.5 & 2.669 & \(1.3 \pm 0.2\) & \(1.3 \pm 0.2\) & ... & D \\ 
10        & 34.0 & 2.7 & 2.668 & \(1.0 \pm 0.2\) & \(1.0 \pm 0.2\) & ... & U \\
11        & 12.7 & \(-12\) & 2.664 & \(1.5 \pm 0.1\) & \(1.5 \pm 0.1\) & ... & M \\
\enddata
\tablecomments{Column 2, 3: Angular separation from the center\(^b\), where +x and +y are in the North and West direction, respectively; Column 5: CO(3–2) line luminosity corrected for magnification; Column 6: Dust mass corrected for magnification; Column 7: Kinematic classification of either Disks (D), Mergers (M) or Undetermined (U)}
\tablenotetext{a}{Measurements made from image c (1c, 4c, 6c)}
\tablenotetext{b}{Center position: RA=131.70877, DEC=15.09936}
\end{deluxetable*}

\subsection{A Diverse Population of Galaxies} \label{Sec:A Diverse Population of Galaxies}
We preface this analysis with the caveat that without reconstruction of the resolved source plane, intrinsic morphologies may not be preserved when high magnification gradients and shearing occur across individual objects. However, in most cases, magnification gradients are minimal, such that their kinematic and morphological features can still be ascertained. By contrast, the images of ID1 are magnified significantly, leaving its kinematic structure highly uncertain. A comprehensive kinematic analysis in the resolved source plane will be conducted in future work. Nonetheless, here we begin to highlight some of the prominent features observed in the J0846.BG protocluster core candidate that reflect the diverse properties of its members.
\par
With the generated moment-1 velocity maps (Section \ref{Sec:The CO Detections at $z=2.67$}), we search for smooth, monotonic velocity gradients, indicative of ordered rotation within gas disks \citep{Rizzo_2023} (Figure \ref{fig:4}). We find that six of the 11 galaxies (ID2, ID3, ID4, ID5, ID7, ID9) are consistent with such properties. Rest-frame UV/optical imaging allows us to examine the distribution of the stellar emission relative to the cold molecular gas and its kinematics, exhibiting a wide-range of properties. This includes members that have their stellar continuum and kinematic major axis well-aligned to offsets in the CO and stellar counterpart with even multiple components (i.e., ID7 host an offset UV-bright star-forming clump detected in r$^{\prime}$ and z$^{\prime}$) or even non-detections (i.e., ID4). Offsets/non detections are expected, given the CO/dust emission represent dusty regions of the ISM \citep[e.g.,][]{Hodge_2012}, where imaging covering wavelengths redder than F160W is necessary to probe the bulk of the stellar continuum embedded within these dusty regions.

In contrast to ordered gas rotation of disks, we also discover evidence of disturbed gas kinematics, characterized by complex morphology with multiple interacting components, and non-monotonic velocity gradients \citep[e.g.,][]{Tadaki_2014, Lee_2019}. We find that two of 11 systems (ID6, ID8) display such iconic signatures of mergers. In both instances, ID6 (ID6.1, ID6.2) and ID8 (ID8.1, ID8.2), consist of two distinct components, with separations between their peaks $\lesssim5$ (physical) kpc apart, with a radial velocity offset of $\lesssim 200$ km s$^{-1}$.
\par
Within ID8, ID8.2 has a FWHM of $\sim 786\pm105$ km s$^{-1}$. That is twice the value for ID8.1, despite being the weaker detection of the two. The line profile of ID8.2 suggests the presence of multiple gas components, requiring a multi-component Gaussian fit (see Figure \ref{fig:2}). Such evidence of disturbed kinematics may indicate fragmented clumps and a turbulent gas reservoir driven by the merger \citep{Mihos_1996, Bournaud_2011,Renaud_2014}. In F160W imaging, ID8.1 has a bright, compact counterpart, while ID8.2 coincides with a more diffuse and faint detection. There is also a tail-like morphology extending to the north (see Figure \ref{fig:4}), further evidence of tidal interactions \citep{Emonts_2015}. The CO(3–2) emission remains fairly compact and does not extend out to these regions, which may indicate that the tidal tail does not comprise of mid-J excited lines and instead requires more diffuse gas tracers (e.g., CO(1–0) or [CI]).
\par
Between ID6.1 and ID6.2 there appears to be a tidal bridge, likely formed due to their gravitational interaction \citep[e.g.,][]{Sparre2022, Lebowitz_2023, Zhong_2024}. It is worth noting that the 6.1c/6.2c images are less distorted due to lower magnification gradient, better retaining their intrinsic morphology, whereas the highly magnified images of 6.1ab/6.2ab better accentuate their distinct sub-structures. For example, only in 6.1ab/6.2ab do we detect a thin optical/IR counterpart, likely originating from star-forming regions able to penetrate through patchy dust in the ISM, further amplified due to lensing \citep[e.g.,][]{Kamieneski_2024}. These magnification factors reach up to $\mu\sim80$ across ID6ab, offering a powerful laboratory to resolve regions at tens of pc scales in future studies.


\section{Discussion} \label{Sec:Discussion}
\subsection{Challenges in Defining J0846.BG as a protocluster core} \label{Sec:Challenges in Defining J0846 as a protocluster core}
The observational definition of protocluster cores and overdensities in general vary in the literature based on their selection method,
often limited in multi-wavelength completeness and areal coverage. Typically, reported DSFG-rich cores have been found to comprise of $4-23$ DSFGs spanning a few hundred kpc, fueling SFRs on the order of 1000s $\textup{ M}_\odot$ yr$^{-1}$ \citep{alberts_2022}. Based on such criteria, we can begin to understand J0846.BG as a protocluster core candidate given its global physical properties. However, we discuss challenges in fully characterizing the structure of this system.


The apparent surveying area of the J0846 field is influenced by gravitational lensing biases. One consequence is the dilation of the image plane, where magnification increases the observed angular sizes, simultaneously reducing the physical area covered in the source plane. The observed angular extent of J0846.BG is $\sim57^{\prime\prime}$, equating to $\sim35^{\prime\prime}$ in the source plane. In the source plane reconstruction (see Figure \ref{fig:8}), the distribution of members suggests the emergence of a coherent structure, extending from the SE towards the NW direction, that may be aligned with a red and blue side in velocity-phase. This is relatively similar to that of the Distant Red Core \citep{Oteo_2018, Long_2020}, appearing to be almost like a filament of galaxies. While this could reflect its true structure, it remains subject to lensing bias as the caustic follows a similar shape. Any structure orthogonal to the caustic is stretched out more (i.e., shear), resulting in the surveying area (source plane) to be smaller in that direction. 


The currently identified members of J0846.BG are limited to those that are sufficiently CO-bright and reside within the $\sim60^{\prime\prime}$ ALMA pointings. Given the limited areal-coverage, further exacerbated by lensing, it is likely there are more DSFGs at $z\sim2.7$. N-body simulations have shown that the progenitors of local massive cluster halos are expected to extend to at least several comoving Mpc at $z > 2$ \citep{Chiang_2013}, far exceeding what is covered in our observations. CO line emitters have been shown to trace these large-scale structures, such as the 46 CO(1–0) sources identified between $z=2.09-2.22$ in the Spiderweb protocluster, over an area of $25$ sq arcmin \citep{Jin_2021}. Thus we caution that the full extent of the J0846.BG could be much larger than the $\sim$arcmin FOV provided by the observations used to conduct this first characterization.

Previous multi-wavelength studies have demonstrated that protoclusters at $z>2$ can host a diverse population of galaxies \citep[e.g.,][]{Umehata_2017,Hill_2020,Champagne_2021}. Unlike DSFGs, sub-mm faint sources would instead trace the more moderately star-forming and less dust-and-gas-rich galaxy populations. In a handful of cases, DSFG-rich protoclusters core have been found to be embedded within larger, $\gtrsim$Mpc scale structure of galaxies \citep[e.g.,][]{Hill_2020,Champagne_2021}. Within the \textit{HST} F160W image's $\sim 2.1^{\prime}\times2.1^{\prime}$ FOV, we photometrically identify an additional $>50$ lensed sources in J0846, that could represent such members also contributing to the total number of galaxies and the full extent of the structure. Follow-up observations are required, as the current imaging and spectroscopic data are insufficient to confirm whether these sources are associated with J0846.BG at $z\sim2.7$.

\subsection{How may this DSFG-rich Core Candidate evolve?}
The growing number of DSFG-rich protocluster cores being identified naturally raises the question of what drives their extreme star formation and how they will continue to evolve. In comparison to large-scale protoclusters spanning tens of cMpc, such protocluster cores only extend up to hundreds of kpc, with co-moving SFRDs more than an order of magnitude \citep{alberts_2022}. These core structures have been shown to host a higher fraction of sub-mm bright galaxies in comparison to their outer regions or in the field \citep[e.g.,][]{Calvi_2023, Araya_2024}. Reproducing such DSFG overdensities may challenge cosmological and galaxy evolution simulations \citep[e.g.,][]{Casey_2016, Hill_2020}, suggesting some special conditions are triggering these extreme environments. 

The virial dynamical mass of J0846.BG could be significantly underestimated (Section \ref{Sec:Dynamical Mass Estimate}), yet we can still explore what this system may evolve into. Based on the evolutionary prediction of \cite{Chiang_2013}, its estimated halo mass of M$_{dyn}=1.2\pm0.6\times10^{13} \textup{ M}_\odot$ may become a $\gtrsim2\times10^{14} \textup{ M}_\odot$ cluster at $z=0$ (and $\sim3\times10^{13} \textup{ M}_\odot$ at $z\sim2$), but the true value could be much larger. We therefore seek to understand the amount of stellar mass that may form from this starbursting system. Assuming a gas-to-stellar mass fraction of unity we estimate a total stellar mass of $M_*\sim2.0\times10^{11}\textup{ M}_\odot$, from our lower limit on the molecular gas mass. We perform a first-order calculation to determine how much more stellar mass will form if the fiducial value of the SFR of $\sim5200  \textup{ M}_\odot$ yr$^{-1}$ is sustained for up to 1 Gyr. We find that J0846.BG could accumulate a total stellar mass of $M_*\sim5\times10^{12}\textup{ M}_\odot$. Assuming a single halo, with a $M_*/M_h=0.03$ \citep{Behroozi_2013} would already yield a halo mass of  $M_h \gtrsim 1\times10^{14}\textup{ M}_\odot$ by $z\sim2$.

The question remains, how long can protocluster cores really sustain their SFR given their molecular gas reservoirs? Violent, gas-rich mergers may serve as a main process for triggering dusty starbursts at $z>2$ \citep{Tacconi_2008, Casey_2016}, where previous work has shown protoclusters to exhibit elevated merger rates \citep{Coogan_2018, Andrews_2024}. 
On the other hand, many $z>1$ DSFGs have been identified to sustain disk rotation \citep{Rizzo_2023}, even in protoclusters \citep{Venkateshwaran_2024, Umehata_2025}, suggesting these disks can survive and/or quickly recover from environmental interactions. Similarly, despite the close proximity of J0846.BG's member, 
we are still find that 6/11 members exhibit kinematic features consistent with ordered-disk rotation (while only two may be mergers and three remain undetermined). In contrast to the short-lived starbursts of mergers \citep[e.g.,][]{Tadaki_2014}, if instead disk are the dominant process driving star formation, its cold mode accretion of gas \citep[e.g.,][]{Dekel_2009} could prevent quenching for longer timescales. 

Often the molecular gas depletion timescale  (\( \tau_{\text{dep}} = \frac{M_{\text{gas}}}{\text{SFR}} \))  is used to determine how long star formation can be sustained given the available molecular gas budgets. The short gas depletion timescale of J0846.BG measured to be $<50$ Myr could be much longer depending on the value of $\alpha_{\text{CO}}$ used, according to the order of magnitude dispersion found in \cite{Harrington2021} for the larger PASSAGES sample (see Section \ref{Sec:First order estimates of molecular gas and dust masses}). 
This is also assuming that no additional molecular gas is being replenished by any large-scale reservoirs stored throughout the circum-galactic medium (CGM) or the proto-intracluster medium (proto-ICM). \cite{Zhou_2025} detected up to 75\% excess of CO flux in SPT2349-56 using Atacama Compact Array observations compared to 12m configuration higher-resolution data, hidden in extended and diffuse gas. 

If J0846.BG can really continue to form a substantial amount of stellar mass, it may represent the progenitor population of massive, $M_{h}>10^{14}\textup{ M}_\odot$ clusters identified at $z\approx1-2$. But even at $z\lesssim2$ there is variation in the observed level of star-forming activity \citep[e.g.,][]{Alberts_2016}. Down to $z\sim1$, clusters have been found to host BCGs with significant star formation \citep{Webb_2015}, and instances of gas-rich members in high-density environments \citep[e.g.,][]{Noble_2017}, indicating they have yet to transition to become like their passive elliptical/lenticular galaxy descendants. 
Conversely, there is evidence of heightened AGN activity \citep{Alberts_2016, Shimakawa_2024} and the presence of massive post-starburst populations with signs of quenching within their environments \citep{Strazzullo_2016,Maier_2019}. The inhomogeneous and limited sample has made it especially difficult to determine how high-redshift clusters are directly linked to their progenitor counterparts. Addressing these open questions is beyond the scope of this discovery paper, where future work on J0846.BG will better constrain the gas-excitation properties and total molecular gas budget to study how different mechanisms drive and eventually suppress star formation. This is crucial to understanding the transition between evolutionary stages of large-scale structure over time.

\section{Conclusion} \label{Sec:Conclusion}
In this work, we have reported the discovery of a protocluster core candidate (J0846.BG) at $z=2.67$, being strongly-lensed by a foreground galaxy cluster (J0846.FG) at $z=0.77$. This \textit{Planck}-selected, strongly-lensed sources is one of the fields within the PASSAGES sample, yet remains the most extreme out of all of the strong lensing fields due to the rare instance of strong lensing one of the most molecular gas and dust-rich protocluster core candidates ever reported. 

We performed the first detailed analyses of the Cycle 5 ALMA Band 3 observations to image the CO(3-2) line and dust continuum emission to spatially resolve the initial detections from the LMT. 
In the first initial characterization of this field we have constructed a lens model to begin to study the physical properties of the member galaxies and the foreground lensing cluster using multi-wavelength observations obtained from the ALMA, VLA, \textit{HST}, Gemini and VLT.





Our main results are summarized as follows:

\begin{itemize}
\item ALMA (0.5$^{\prime\prime}$) observations uncover 18 CO(3–2) line detections between $z=2.660-2.669$ being lensed by a foreground cluster. The foreground cluster was found to be at $z=0.77$ with an ensemble of spectroscopic data obtained from Gemini/GMOS, VLT/MUSE and VLT/FORS2. 

\item We construct a gravitational lens model of the J0846.FG foreground cluster, via a parametric approach. We report that J0846.BG consists of 11 galaxies that are lensed into 18 unique CO(3–2) line detections, spanning a physical extent in the reconstructed source plane image of $280 \times 150$ kpc and within a velocity range of $\Delta V\approx800$ km s$^{-1}$ with a velocity dispersion of $\sigma_{v}=246\pm72$ km s$^{-1}$.

\item  J0846.BG is one of the most actively star-forming protocluster core candidates to ever be reported. We use the newly constructed lens model to measure the total intrinsic global star formation rate, corrected for magnification of \(\mathrm{SFR}=5200^{+3200}_{-2000}~\mathrm{M}_{\odot}~\mathrm{yr}^{-1}\). We estimate the dynamical mass by applying the virial theorem to the measured galaxy velocities, finding a value of M$_{dyn}=1.2\pm0.6\times10^{13} \textup{ M}_\odot$. 
\item The strongly lensed background galaxies are a diverse population, with evidence of well-ordered disk rotation (6/11) and disturbed kinematics driven by mergers/tidal interactions (2/11).
\end{itemize}
J0846 offers a remarkable laboratory to study the driving mechanisms of star formation within extragalactic protocluster environments -- the most massive progenitors to modern day large-scale structures and galaxy clusters. Given the benefits of strong gravitational lensing, physical scales can be resolved at exquisite detail to better understand how this system of at least 11 dusty star-forming galaxies will evolve under this assembly within 300 kpc. The discovery of this field now enables a detailed investigation to answer some of the outstanding questions in the field of galaxy evolution and protocluster formation in the early Universe. It is still unclear how galaxy environments may affect star formation within individual systems, nor is it obvious how gas is accreted onto actively forming galaxies. It is therefore necessary, for example, to explore resolved multi-wavelength diagnostics characterize the thermal/ionization gas states that may exist between systems and to account for the bulk cooling energy present to explain its subsequent collapse into a local massive galaxy cluster. These 11 dusty, CO-emitting sources are all found to be within 300 kpc, suggesting a wide range of interactions that could lead to tidal interactions and ram-pressure stripping and AGN feedback, where strong lensing enables a magnified view into these physical processes at high redshift. Further spatially unresolved investigation of the full CO/[CI] line SED began with the IRAM-30m/EMIR and APEX/PI230+FLASH \citep{Harrington2021}, which has led to a series of VLA and ALMA programs of the field using the Atacama Compact Array and 12-m array (ALMA: program 2023.1.00299.S, PI: N. Foo, program 2022.1.01311.S, PI: P. 
Kamieneski, programs 2022.1.01282.S, 2021.2.00088.S., PI K. Harrington, VLA: program 25A-310, PI: N. Foo). Such observations will enable analyses of the large-scale environment and spatially resolved multi-line CO/[CI] gas excitation and kinematic properties within J0846 (Harrington in prep; Foo in prep.) to place the individual objects along the main sequence for star-forming galaxies, derive the molecular gas to stellar mass fractions and determine the available molecular gas content for sustaining the ongoing star formation to gain further insight into the underlying mechanisms that drive and subsequently quench these starburst events. 

\acknowledgments
We thank the anonymous referee for helpful comments and suggestions that improved the manuscript. NF was supported in part off of the Arizona NASA Space Grant Consortium, Cooperative Agreement 80NSSC20M0041. N.F. and K.H. would like to thank the European Southern Observatory Office for Science in both Vitacura, Chile and Garching, Germany for the Science Support Discretionary Fund that enabled NF to visit Chile and focus on this project. KH and NF would like to thank Manuel Aravena, Jorge Gonzalez-Lopez, Manuel Solimano and others at the Universidad Diego Portales in Santiago, Chile for useful discussions and perspectives. N.F. would like to thank the National Radio Astronomy Obsevatory (NRAO) for funding through the Student Observing Support program. E.F.-J.A. acknowledge support from UNAM-PAPIIT projects IA102023 and IA104725, and from CONAHCyT Ciencia de Frontera project ID: CF-2023-I-506. A.N. acknowledges support from the National Science Foundation through grant AST-2307877. AZ acknowledges support by Grant No. 2020750 from the United States-Israel Binational Science Foundation (BSF) and Grant No. 2109066 from the United States National Science Foundation (NSF); and by the Israel Science Foundation Grant No. 864/23. A.D. acknowledges support from project PID2022-141915NB-C22/MCIN/AEI/10.13039/501100011033. RAW, and SHC acknowledge support from
NASA JWST Interdisciplinary Scientist grants NAG5-12460, NNX14AN10G and 80NSSC18K0200 from GSFC. This paper makes use of the following ALMA data: ADS/JAO.ALMA 2017.1.01214.S. ALMA is a partnership of ESO (representing its member states), NSF (USA) and NINS (Japan), together with NRC (Canada), NSTC and ASIAA (Taiwan), and KASI (Republic of Korea), in cooperation with the Republic of Chile. The Joint ALMA Observatory is operated by ESO, AUI/NRAO and NAOJ. This work is also based on observations made with the NASA/ESA Hubble Space Telescope. The data were obtained from the Barbara A. Mikulski Archive for Space Telescopes (MAST) at the STScI, which is operated
by the Association of Universities for Research in Astronomy (AURA) Inc., under NASA contract NAS 5-26555 for \textit{HST}. The specific observations analyzed can be accessed via\dataset[DOI: 10.17909/vxg2-cn33]{http://doi.org/10.17909/vxg2-cn33
}. Facilities: ALMA, \textit{HST}, VLA, MAST, Gemini, VLT.

\bibliography{J08/J08_ref}{}
\bibliographystyle{aasjournal}

\appendix
\restartappendixnumbering


\section{Further Details on Data Reduction and Analysis}
\subsection{VLA: 6 GHz Continuum Imaging}
Photometry for the 6 GHz imaging is performed using source extraction software BLOBCAT \citep{Hales_2012} which utilizes a flood-filling algorithm to detect blobs of pixels to identify candidate sources. We report that five of the CO(3–2) sources in J0846.BG have corresponding 6 GHz emission detections at $\geq3\sigma$ significance (Table \ref{Table:2}). Furthermore, we detect two radio counterparts to cluster member galaxies in J0846.FG, one of which exhibiting a wide ``C'' shape morphology (see Section \ref{Sec:A Radio Wide Angle Tail Galaxy}).

\subsection{Optical-Infrared Imaging} \label{Appendix:Optical-Infrared Imaging} 
Photometry was obtained from the F160W, r$^{\prime}$ and z$^{\prime}$ band imaging utilizing \texttt{SExtractor}, applying nominal settings. We generate matched r$^{\prime}$ and z$^{\prime}$ catalogs utilizing the z$^{\prime}$ band as the detection image for aperture extraction to measure photometric colors for foreground cluster member identification. To measure their zero-point AB magnitudes for r$^{\prime}$ and z$^{\prime}$ we calibrate to Sloan Digital Sky Survey (SDSS) photometry by cross-matching objects in the field. We compute limiting magnitudes by determining the drop-off in object counts measuring values of $m_{\text{z$^{\prime}$}}\approx 24.3$ mag and $m_{\text{r$^{\prime}$}}\approx 24.8$ mag. F160W photometry for any counterpart detections of the 18 CO images is provided in Table \ref{Table:2}. The 18 images are undetected in the Gemini photometry except for ID7 ($m_{\text{r$^{\prime}$}}= 24.55$ mag, $m_{\text{z$^{\prime}$}}=23.45$ mag) and ID8.1 ($m_{\text{z$^{\prime}$}}=24.13$ mag).

\subsection{Optical Spectroscopy} \label{Appendix:Optical Spectroscopy} 
We compile the full catalog of redshifts extracted from the optical spectroscopy across the GMOS, FORS2/VLT and MUSE/VLT observations. In total, we obtain spectra of 95 objects, and are able extract secure redshifts from 29 of them. Three of the sources have measurements from both FORS2 and MUSE while one of them also has a GMOS redshift value, yielding a total of unique 25 sources. We measure consistent redshift in all of the duplicate spectra, across the different instruments. To substantiate a secure redshift measurement we require at least two spectral features, specifically absorption/emission lines and consistent stellar continuum properties. Given the wavelength coverage, typical lines identified include strong absorption metal lines including Ca H \& K, G-band, and Mg I $\lambda$ 5173. These lines were consistently identified for cluster member galaxies in J0846.FG. Additionally we identify expected continuum features (i.e., 4000$\text{\AA}$ break) to further support our redshift determinations. For a handful of objects we also detect strong emission line features including the Balmer line series ($\text{H}\alpha$ through $\text{H}\delta$),  [N II] $\lambda\lambda$6548, 6583, [S II] $\lambda\lambda$6716, 6731, [O III] $\lambda\lambda$4959, 5008, [O II] $\lambda\lambda$3727, 3729. The full catalog can be found in Table \ref{Table:A1} of the Appendix. No optical spectroscopic measurement was able to be obtained for any of the background multiply-imaged systems. The geometric lens model predicted redshifts and their magnification factors of the non-CO detected image systems are shown in Table \ref{Table:A2}. 

\begin{deluxetable}{ccccccccc}[h]
\label{Table:A1}
\tablecaption{Optical Spectroscopic Redshifts}
\tablecolumns{9}
\tablewidth{0pc}
\tablehead{
\colhead{ID} & \colhead{R.A.} & \colhead{Decl.} & \colhead{$z_{spec}$}& \colhead{Instrument}\\
}
\startdata
\ 
962  & 131.708768 & 15.099363 & 0.357 & M\\ 
1396  & 131.703667 & 15.102103 & 0.750 & M+F+G\\ 
1293 & 131.706096 & 15.101416 & 0.755 & M\\ 
645 & 131.714097 & 15.097136 & 0.768 & M\\ 
761  & 131.708501 & 15.096806 & 0.768 & M+F\\ 
1574  & 131.719610 & 15.094445 & 0.381 & M\\ 
1715  & 131.715481 & 15.095193 & 0.399 & M\\ 
612  & 131.710239 & 15.093068 & 0.770 & M+F\\ 
456  & 131.715004 & 15.092078 & 0.768 & M\\ 
879  & 131.702744 & 15.098780 & 0.758 & F\\
1   &131.709394  & 15.081012  & 0.762 &F\\
400 & 131.718162  & 15.091785  & 0.772 & F\\
11 &  131.702195 &  15.143417 &   0.3660 & G\\
1208 & 131.705526  &15.117846 & 0.3675  & G\\
 2344 & 131.667931 & 15.090582 & 0.5950  & G\\
1036 & 131.698748 & 15.121020 & 0.7060  & G\\
23 & 131.703237 & 15.138873 & -0.006 & G \\
87 & 131.692283 & 15.123852 & -0.01 & G \\
1511 & 131.688581 & 15.110132 & -0.005 & G \\
1612 & 131.716294 & 15.109275 & -0.006 & G \\
2825 & 131.693097 & 15.086560 & 0.067 & G \\
3932& 131.704219 & 15.060462 & -0.007 & G \\
4033& 131.700998 & 15.058060 & -0.007 & G \\
4134& 131.699710 & 15.055824 & -0.099 & G \\
55 & 131.744329 & 15.130727 & 1.548 & G \\
\hline
\enddata
\tablecomments{ Column 1: ID; Column 2: R.~A.; Column 3: Decl.; Column 4: Spectroscopic Redshift; Column 5: Instrument used to obtain spectra where M, F and G represent MUSE, FORS2, and GMOS, respectively.}
\end{deluxetable}

\begin{deluxetable}{ccccccccc}[h] \label{Table:A2}
\tablecaption{Non-CO detected Multiply-Imaged Systems}
\tablecolumns{9}
\tablewidth{0pc}
\tablehead{
\colhead{ID} & \colhead{R.A.} & \colhead{Decl.} & \colhead{$z_{geo}$}& \colhead{$\mu$}\\
}
\startdata
\ 
12a & 8:46:48.6732 & +15:06:00.863& \(2.90 \pm 0.08\) & \(5.7 \pm 2.0\)\\ 
12b & 8:46:48.5340 & +15:05:49.363&...& \(3.1\pm 0.3\)\\ 
12c & 8:46:48.5138 & +15:05:55.978&...& \(2.3 \pm 0.3\)\\ 
12d & 8:46:48.7477 & +15:05:55.123&...& \(2.0 \pm 1\)\\ 
13a & 8:46:48.5295 & +15:05:57.771&\(2.77 \pm 0.08\) &\(9.3 \pm 6.2\)\\ 
13b & 8:46:48.5867 & +15:05:59.036&...& \(4.2 \pm 6.1\)\\ 
14a & 8:46:48.3953 & +15:06:00.559& \(2.84 \pm 0.08\) & \(6.4 \pm 1.1\)\\ 
14b & 8:46:48.3595 & +15:05:57.573&...& \(4.2 \pm 1.1\)\\ 
\hline
\enddata

  \label{tab_GGs}
\end{deluxetable}
Notably, all of our optical spectroscopic observations were obtained via filler time programs, for which optimal weather conditions and good seeing were not guaranteed. Moreover, the MUSE observations consisted of a single pointing that only partially covered the central region of the HST FOV, while the wider-field GMOS and FORS2 multi-object slit spectroscopy only sparsely covered the set of cluster members. 

\section{More Details on the Foreground Cluster}

\subsection{The Perturber: A Bright Elliptical}\label{Sec:The Perturber: A Bright Elliptical}
The brightest galaxy identified in the F160W image is an elliptical located conspicuously near the center of the field (see Figure \ref{fig:4}), and was initially mistaken to be the BCG of the cluster. However, subsequent MUSE spectroscopy revealed that this galaxy is at the redshift of $z=0.357$, directly along the line of sight to J0846.FG at $z=0.77$. Its optical/IR color is also different compared to the other confirmed cluster members (see Figure \ref{fig:7}). Although this galaxy is physically unrelated to the cluster structure, it may still have non-negligible contribution towards the overall lensing potential. We gain a prior on its dynamical mass by measuring line widths of its spectral features in its 1-D MUSE spectra where $\text{FWHM}=2\sqrt{2\text{ ln}2} \sigma$, after correcting for the instrumental resolution. Interestingly, there are a handful of spectroscopically confirmed galaxies at $z=0.3-0.4$ scattered throughout the field, but their redshift differences are too large to verify any significant structure.

\subsection{Dynamical Mass Estimate} 
The velocity dispersion is calculated based on their proper velocities using the biweight sample variance, yielding $\sigma_{v}=1446\pm471$ km s$^{-1}$. 
To estimate the dynamical mass of this core structure, we employ the virial theorem assuming a spherical mass distribution (Equation \ref{Eq:Dyn Mass}).
R, is the maximum physical separation over all members, measured to be $400$ kpc. Here we estimate the dynamical mass to be M$_{dyn}=5.8\pm3.8\times10^{14} \textup{ M}_\odot$.

\subsection{A Radio Wide Angle Tail Galaxy} \label{Sec:A Radio Wide Angle Tail Galaxy}
In the VLA 6 GHz continuum observations, we detect two spectroscopically confirmed members in J0846.FG. 
One of these detections exhibits radio tails bent back into a ``C''-shape morphology (see top-right inset in Figure \ref{fig:3}), identified to be radio Wide-Angle Tails (WAT) \citep{Owen_1976}. It is believed that the relative motion of a galaxy moving through the intracluster medium (ICM) magnetic field can trigger radio jets that get swept back into tails \citep{O'Donoghue1993}. WATs are commonly associated with the Brightest Cluster Galaxy (BCG) of a cluster \citep[e.g.,][]{Owen_1976, Simon1978}, and are predominantly found in merging systems \citep{O'Dea_2023}. Moreover, clusters undergoing a merger can produce large scale ($\sim 100$ kpc) bulk velocities in the ICM, i.e., sloshing which intensifies the ram pressure \citep{Markevitch2007}, where BCGs in non-relaxed systems can be found away from the cluster center, with larger velocities ($\sim200-400$ km s$^{-1}$) relative the cluster \cite[e.g.,][]{Beers_1991}. With current datasets there is no clear identification of a BCG in J0846.FG, which may also indicate a merger scenario \citep[e.g.,][]{Okabe_2019}. Regardless, this radio WAT galaxy serves as the tentative BCG as it is the brightest confirmed member, albeit only about $\Delta m_{\text{F160W}}\sim0.1$ mag brighter then the next brightest cluster galaxy at $z=0.77$. Additionally, this galaxy along with one other member is detected in 3mm continuum. 
Interestingly, the brightest galaxy in the \textit{HST} FOV is in the foreground of the cluster (see Section \ref{Sec:The Perturber: A Bright Elliptical}).

\section{Global Modified Blackbody Fitting to the Dust SED} \label{Appendix:Global Modified Blackbody Dust SED}
\begin{figure}
\centering\includegraphics[scale =0.65]{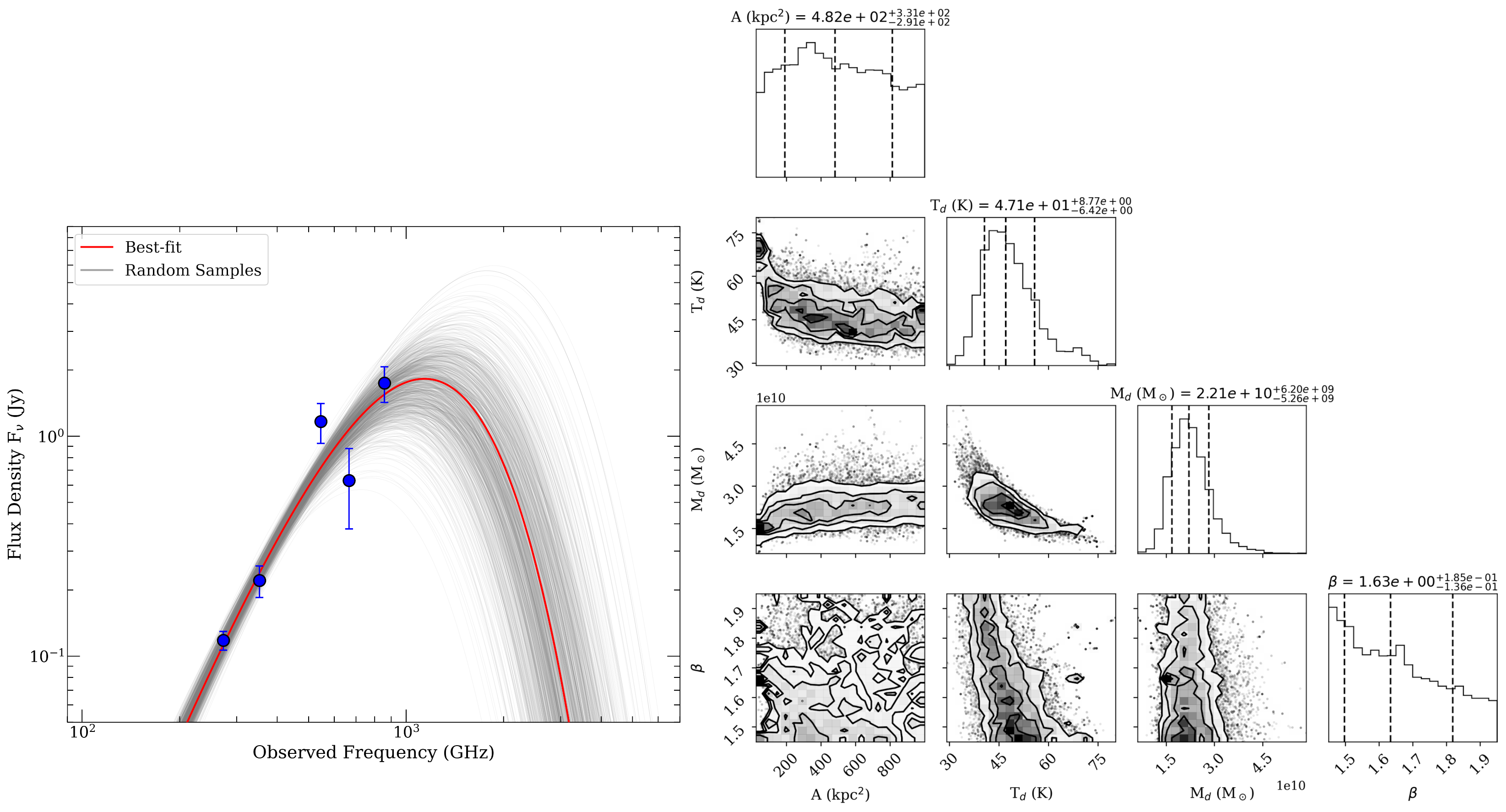} 
\caption{Left: The global modified blackbody dust SED model fit for J0846.BG, utilizing the unresolved dust continuum measurements \citep[from][]{Harrington2021, Berman2022}. We plot the best fit model (red curve) and 1000 randomly sampled models from the MCMC posterior (gray curves), as well as the photometric data points (blue points). Right: Corner plot displaying the posterior likelihood distribution for the parameter space of the modified blackbody model fit. The dashed lines on the 1D posterior distribution represent the 16th, 50th, and 84th percentile.
} 
\label{Table:C1}
\end{figure}

We model the dust SED by fitting the measured flux densities from spatially unresolved observations of the J0846.BG system using the following equation for a modified blackbody \citep[see e.g.][]{DaCunha2013, Fernandez_Aranda_2025}: 
\begin{equation}
S_{\nu_{\text{obs}}} = \frac{A}{D_L^2} [B_{\nu_{\text{rest}}}(T_d) - B_{\nu_{\text{rest}}}(T_{\text{CMB}})][1 - e^{-\tau_{\nu_{\text{rest}}}}  ](1 + z)
\end{equation}
giving the flux density, $S_{\nu_{\text{obs}}}$, observed at frequency $\nu_{\text{obs}}$, where $B_{\nu}$ is the Planck function for a given temperature, $A$ is the physical area of the emission, $D_L$ is the luminosity distance, $T_d$ is the dust temperature, $T_{\text{CMB}}$ is the temperature of cosmic microwave background (CMB) radiation at the source redshift, $z$. The optical depth of the dust, $\tau_{\nu}$ is defined as

\begin{equation}
\tau_\nu = \kappa_0 \left(\frac{\nu}{\nu_0}\right)^\beta \frac{M_d}{A}
\end{equation}
where $\beta$ is the dust emissivity index, $M_d$ is the dust mass, and $\kappa_0$ is the dust mass absorption coefficient at a reference frequency $\nu_0$. We adopt a value of $\kappa_0 = \kappa_{850\mu\text{m}} = 0.077\, \text{kg}^{-1}\text{m}^2$ where $\nu_0 = 352.7\, \text{GHz}$ ($850\,\mu\text{m}$) following \cite{Dunne_2000} for their sample of SMGs. We use a MCMC approach with the \texttt{emcee} \citep{Foreman-Mackey_2013} Python package to sample the posterior distribution, setting $A$, $T_d$, $M_d$, and $\beta$ as free parameters. Fitting to the unresolved dust continuum measurements of the entire system \citep[from][]{Harrington2021, Berman2022}, we find $A=482^{+331}_{-291}$ kpc$^{2}$, $T_d=47.1^{+8.8}_{-6.4}$ K, $M_d=2.21^{+0.62}_{-0.53}\times10^{10}\textup{ M}_\odot$, and $\beta=1.63^{+0.19}_{-0.14}$. We show the modified blackbody SED fit and the corresponding corner plot for the posterior distribution of the parameter space in Figure \ref{Table:C1}. We determine the total apparent IR luminosity by integrating 
the SED fit from $8-1000\mu$m, finding a value of \(\mu\mathrm{L}_{\mathrm{IR}} = 3.75^{+2.16}_{-1.21} \times 10^{14}~\textup{L}_{\odot}\).

\end{document}